# AI-Based Detection, Classification and Prediction/Prognosis in Medical Imaging: Towards Radiophenomics

Fereshteh Yousefirizi[1*], Pierre Decazes[2], Amine Amyar[3], Su Ruan[3], Babak Saboury[4,5,6], Arman Rahmim[7,8,9]

[1]Department of Integrative Oncology, BC Cancer Research Institute, Vancouver, BC, Canada
[2] Department of Nuclear Medicine, Henri Becquerel Cancer Center, Rouen, France
[3] QuantIF-LITIS EA 4108, University of Rouen, Rouen, France
[4]Department of Radiology and Imaging Sciences, Clinical Center, National Institutes of Health, Bethesda, MD, USA
[5]Department of Computer Science and Electrical Engineering, University of Maryland Baltimore County, Baltimore, MD, USA,
[6]Department of Radiology, Hospital of the University of Pennsylvania, Philadelphia, PA, USA
[7]Departments of Radiology and Physics, University of British Columbia
[8]Senior Scientist & Provincial Medical Imaging Physicist, BC Cancer
[9]BC Cancer Research Institute

Corresponding author: frizi@bccrc.ca

**Synopsis:** Artificial intelligence (AI) techniques have significant potential to enable effective, robust and automated image phenotyping including identification of subtle patterns. AI-based detection searches the image space to find the regions of interest based on patterns and features. There is a spectrum of tumor histologies from benign to malignant that can be identified by AI-based classification approaches using image features. The extraction of minable information from images gives way to the field of radiomics and can be explored via explicit (handcrafted/engineered) and deep radiomics frameworks. Radiomics analysis has the potential to be utilized as a noninvasive technique for the accurate characterization of tumors to improve diagnosis and treatment monitoring. This work reviews AI-based techniques, with a special focus on oncological PET and PET/CT imaging, for different detection, classification and prediction/prognosis tasks. We also discuss needed efforts to enable the translation of AI techniques to routine clinical workflows, and potential improvements and complementary techniques such as the use of natural language processing on electronic health records and neuro-symbolic AI techniques.

**Key Points:**
- Artificial intelligence (AI) techniques are being increasingly explored in medical imaging. Innovations in machine learning (ML) and deep learning (DL) have helped unlock potentials of AI for successful applications.
- Patient health information, including demographic information, electronic medical record notes, diagnostic imaging at different time-points, and radiologist reports, along with radiomic features can be used as input to AI techniques for detection, classification and outcome prediction.
- There is significant value for reliable and automated AI-based tools for improved clinical task performance.
- We also discuss needed efforts to enable translation of AI techniques to routine clinical workflows, and potential improvements and complementary techniques such as the use of natural language processing on electronic health records and neuro-symbolic AI techniques.
- The phenomics approach as it is introduced for precision medicine enable the systematic discovery of structural and functional patterns associated with disease presentation or drug response, we considered the application of AI techniques for radiophenomics i.e. radiomics and phenomics in this paper.

**Disclosure Statement:** Authors do not have anything to disclose regarding conflict of interest with respect to this manuscript.

## 1. Introduction

The task of clinical interpretation of medical images starts with the scanning of the presented image to detect the suspicious finding ("observation" in RadLex terminology (RID5) [1] which is also used in various Reporting and Data Systems (RADS), such as LI-RADS for liver imaging [2]), with subsequent attention to details (characterize) in combination with clinical history (contextualization) in order to classify the finding/observation as "normal variant", benign or malignant with certain level of confidence and various amount of details (diagnosis). There are two kinds of errors in this process: "perceptual error" (not detecting the finding) and "faulty reasoning" (attributing the discovered observation to wrong cause) [3]. Using Kahneman terminology [4], these two categories are the errors of "system 1" thinking (automatic and intuitive) versus "system 2" thinking (logical and based on reasoning).

Analysis of medical errors in diagnostic radiology reveals 71% of all errors are "missed findings" [5]. Kim-Mansfield radiologic error classification of these "perceptual errors" demonstrates two sub-categories: detection error (looking but not seeing, 60%) and suboptimal/insufficient scanning (40%, either due to satisfaction of search or locations outside the intended focus of image, such as a pulmonary nodule in CT of Abdomen). This high rate of error shows the complexity of "detection process" even for human brain.

There are multiple sources of lexical ambiguity here. The terminology of visual processing in neuroscience is based on biology of the brain and connected networks. Those terms might be used in completely different sense in computer vision and image processing field. It is important to be cognizant of this potential confusion. The brain visual processing pathways are schematically shown in Figure 1(two parallel streams of "what pathway" [involved in recognition, identification, and categorization of visual stimuli] and "where pathway" [involved in spatial attention] make the process more complex [6]). In neuroscience, vision has three stages: encoding, selection, and decoding/inference. Encoding is to sample and represent the visual input. Selection or "attention selection" is to select tiny fraction of input for further analysis; and decoding is "recognition" of the object [6, 7]. Perception means "inferring or decoding the visual scene properties" from the visual input (to form percept). Object recognition (OR) refers to the ability to identify the object in the scene by matching the processed input (percept) with "mental representation" of the object. OR is hierarchical and complex [8-11]. Failure in recognition of an object is called *visual agnosia* and has two subcategories: *apperceptive agnosia* (failure in recognition due to perception error; inadequate integration of simple sensory information such as color or edges to form an integrated property; e.g. one cannot identify the shape) versus *associative agnosia* (perception is intact but one still cannot recognize the objects in general [*object agnosia*] or specific objects such as face [*prosopagnosia*] (face blindness), word [*agnostic alexia*], or location/environment [*topographagnosia*], etc).

In computer vision, object detection is defined as the combination of localization (identifying the object location) and some level of classification (identifying a broad object category; e.g. when applied to medical imaging: "likely not benign", thus eliminating some normal physiologic uptake patterns). Both localization and classification tasks need robust features; however, they have different properties. The shift-variant task of localization often searches the entire image domain and is cognizant of spatial coordinates, while classification is a shift-invariant task that mainly requires features from the zoomed-in part of the image to determine the object category [12].

Artificial intelligence (AI) based detection algorithms have been utilized in diagnostic radiology with various intentions: to accelerate detection of critical findings (such as automatic detection of large vessel occlusion detection [13] for stroke management), to facilitate the processes with high cognitive burden (such as detection of lung nodule [14] ), to augment the diagnostic accuracy of radiologist (such as detection of suspicious lesion in mammography [15] [15]), to prioritize the reading list [16] , and many more use cases. It is important to distinguish between "detection" and "diagnosis". Occlusion of large pulmonary vessel is a "detection" but identifying that finding as pulmonary thromboembolism is a "diagnosis". This difference is not merely semantic and has significant legal and regulatory consequences (c.f. Computer Aided Detection, CADe [17], is regulated under 21 CFR 892.2050 and 2070; however Computer Aided Diagnosis, CADx, is regulated under 21 CFR 892.2060 [18]). These implications have an important role in "terminology confusion" of scientific field. A task could be very similar to diagnosis but the claim may be merely detection and results communicated using different terms depends of the audiences, context and intention of the writer.

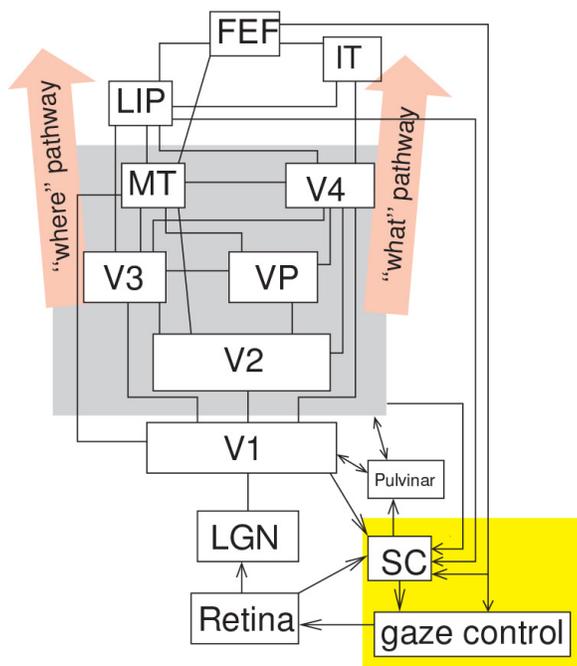

Figure 1: Neuroscience of visual processing demonstrating the hierarchy of information processing from lower stage (retina) to higher stage (FEF). Each neuron typically responds to inputs from limited extend of visual space called its receptive field (RF). The RF of retinal neurons are small (0.06 degree) and gets progressively larger (V4 neuron: 10 degree; IT neuron: 20-50 degree). In early areas such as retina RF and V1 RF are fixed and independent of "attention" while it is variable in later areas (consider range of RF in IT). Each area has millions of neurons (about half of areas in monkey brain are involved in vision). LGN: Lateral geniculate Nucleus (part of thalamus). V1: visual area one in occipital cortex; largest visual cortex in brain. V2: visual area 2. V4: visual area4. MT: middle temporal area (terminology from macaque); neurons are sensitive to motion. LIP: lateral intra-parietal area; involved in decision making for eye movement. IT: infratemporal cortex that responds to complex shapes. FEF: frontal eye field in frontal cortex. SC: superior colliculus (part of midbrain); both FEF and SC are involved in the control of eye movement .(from [6] with permission)

There are four sources of "terminology confusion": 1. difference between visual neuroscience and AI-based computer vision; 2. evolving practice and developments in medical image analysis and inconsistency of usage among scientific groups; 3. borrowing terms from other fields; e.g. signal detection theory or psychology, and 4. legal implications of specific terms. To make this "tower of babel" a better place, we have to redefine these words/terms in this paper in a careful way to decrease the ambiguity and increase the clarity of the language.

From an AI point of view, detection and characterization both involve classification tasks (broad definition). Here, we refer to "detection" as a task that finds a subspace in image scene (localization) that contains a "specific object class" (classification) with a certainty level. However, we use a narrower definition of "classification" compared to statistical/AI terminology. Here, "classification" refers to the process of *characterization and identification* of already-detected/localized lesions. It can be an end-to-end process (input: image; output: category) or composite process (combination of *feature extraction* plus *statistical classification* of the features). However, exam (image) classification can also occur in the context of prediction/prognosis, enabling risk stratification, which we additionally review in this work. Stratification of cancer into reliably distinct risk subgroups enables personalization of treatment. Radiomics is the large numbers of quantitative features combined with machine learning methods to determine relationships between the image and relevant clinical outcomes[19]. We can call the process of "categorization of features into clinically meaningful phenotypes" phenomics [20-26]. If the task of "image classification" has significant clinical value (diagnostic or prognostic), we call it radiophenomics. It could be either an end-to-end process (input: image; output: category/phenotype) or composite process (combination of radiomics and phenomics).

In this work, we first discuss advanced image quantification, leading to the field of radiomics in general, including *explicit* (handcrafted/engineered) radiomics vs. *deep* radiomics paradigms. Next, our focus will be specifically on (i) detection, (ii) classification (characterization) of detected lesions, and (iii) prediction/prognosis tasks. Unlike (ii), (iii) usually involve utilization of a wide search space (e.g. various identified tumors and areas just outside, and distances between them, etc.). As an example, we describe current status on different AI applications in PET imaging for detection, classification, and prognosis/prediction, and share our vista about upcoming opportunities and future directions while reminding potential hurdles in the path of translating these methods into clinical practice. The examples of PET imaging in this article is limited to cancer imaging by PET and here we do *not* review AI techniques used in non-oncological applications; e.g. cardiovascular SPECT/PET and brain PET [27].

## 2. Advanced Image Quantification; the Field of Radiomics

Medical images contain significant minable and potentially valuable quantitative information beyond what is nowadays captured in routine clinical evaluations, motivating the field of radiomics [28, 29]. An array of AI techniques in the field of medical imaging have emerged in the past decade to derive imaging biomarkers based on this information, utilizing explicit (i.e. handcrafted/engineered) radiomics features or deep radiomics features (i.e. derived via deep neural networks). These techniques have significant potential given their ability to reproducibly extract valuable information, including some that are beyond visual limits. Radiomic features are useful in numerous critical tasks from automated tumor detection in routine screening to Radiogenomics including, classification of patients and the spectrum of tumor histologies, severity scoring, prognosis, clinical outcome prediction (based on clinical and radiological features), treatment planning, and assessment and monitoring response to therapy [30-34].

Figure 2 shows a spectrum of AI applications (including machine learning (ML) and deep learning (DL)) in medical imaging based on varying complexity, expertise, and data. Ongoing studies (reviewed in this work) offer a glimpse into how AI can effectively support radiologists and nuclear medicine physicians by automating certain time-consuming measurements in medical imaging, consequently facilitating improved clinical tasks. Having access to appropriate labeled data with consistent labels is crucial for supervised AI techniques for classification and detection. At the same time, image labels for detection tasks including annotations of tumors location and the relevant classes are not yet extensively available in the field of medical imaging. Weakly supervised techniques have been introduced to tackle this limitation by decreasing the dependency on precise annotations e.g. utilizing image-level annotations [35, 36] for detection and classification tasks. We will review these considerations.

Radiomics is an umbrella term that includes the use of single- or hybrid-imaging modalities, with the potential to identify novel imaging biomarkers for improved detection, classification, staging, prognosis, prediction and treatment planning in different cancers [37-43]. Next, we discuss applications of AI techniques within explicit (handcrafted) vs. deep radiomics paradigms toward derivation and validation of radiomics signatures.

## 3. Radiomics Signatures

Radiomics analyses can augment visual assessments made by radiologists [44]. AI techniques can perform quantitative high-throughput image phenotyping (extracting numerous image-based features) and identifying important discriminative features that individually or in combination form an effective radiomics signature for a given task. The field of radiomics has been introduced and elaborated in an accompanying chapter by Orlhac et al. [29]. The challenging task of detecting suspicious regions and differentiating (classifying) benign vs. malignant nodules may be enhanced using AI techniques that have the potential to "see" beyond human perception using high dimensional data by radiomics, and/or enable potentially more robust clinical task performances.

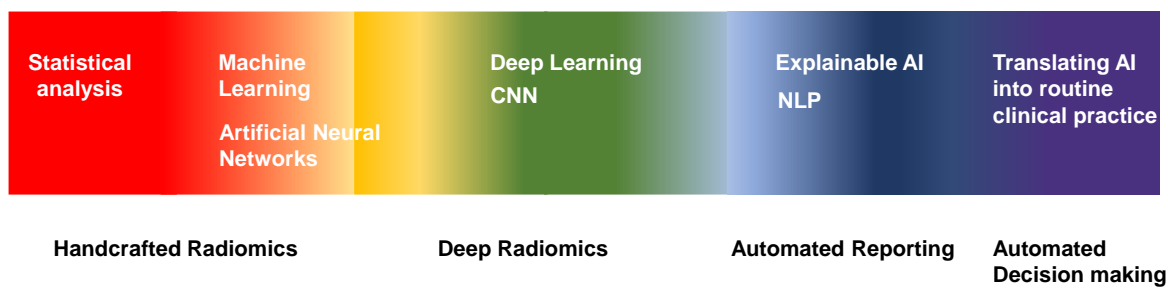

Figure 2: Spectrum of AI applications in medical imaging based on varying complexity, expertise, and data. From the infra-zone (applications of conventional statistical analysis, artificial neural networks (ANNs) and ML to ultra-zone (AI techniques for decision-making and clinical reporting) the complexity of AI applications are amplified. CNNs for different tasks (e.g. detection, classification, prediction/prognosis) are in the middle parts of the spectrum (Adapted from [45] with modifications)

Radiomics signatures have been effectively utilized for detection, classification and prediction/prognosis (e.g. [46, 47]). As specific examples, Lartizien et al. [48] showed that textural features of PET and CT images have a high diagnostic ability in discriminating lymphomatous disease sites from physiologic uptake sites and inflammatory non-lymphomatous. They used ML techniques (SVM and random decision forest) for supervised classification. Nevertheless, the repeatability and reproducibility of textural features should be carefully considered in clinical settings [49]. Kebir et al. [50] assessed the diagnostic value of textural features of PET images for detecting true tumor progression. They found that clustering-based analysis of the heterogeneity features can be used to differentiate the disease progression from pseudo-progression. The present section elaborates some important aspects of radiomics analyses, as these issues are highly relevant and important in specific clinical tasks. Figure 3 depicts two different paradigms of AI-based analysis of medical images, namely explicit vs. deep features: i.e use of ML techniques that utilize explicit features as extracted from segmented images vs. DL techniques that use deep feature extraction or end-to-end learning from images. AI (ML and DL) techniques as mentioned previously can be supervised and weakly/semi/unsupervised, which we have discussed in accompanying chapter [51].Next, we elaborate the explicit (handcrafted/engineered) radiomics vs. deep radiomics paradigms:

**i) Explicit (i.e. handcrafted/engineered) radiomics:** This framework relies on extracting pre-defined with explicit mathematical definitions(e.g. sphericity). Subsequently, statistical or ML-based techniques can be utilized to combine these features into a radiomics signature. ML-based techniques include unsupervised techniques (such as k-means clustering and hierarchical clustering) and supervised techniques (such as logistic regression, decision tree, SVM, etc.). An explicit radiomics workflow, as shown in Figure 4, typically involves the following steps [29, 52]: 1) study design, 2) image acquisition and reconstruction, 3) ROI (VOI) segmentation, 4) spatial resampling and intensity discretization (except for shape features), 5) feature extraction, (6) radiomics model building (using statistical and ML methods), followed by (7) model evaluation, and (8) sharing and reporting of findings. There are a number of tools to extract explicit radiomics features (intensity, shape and texture features), e.g. PyRadiomics [53], LIFEx [54], SERA [55], CERR [56], and others that have been standardized through the Image Biomarker Standardization Initiate (IBSI) [57].

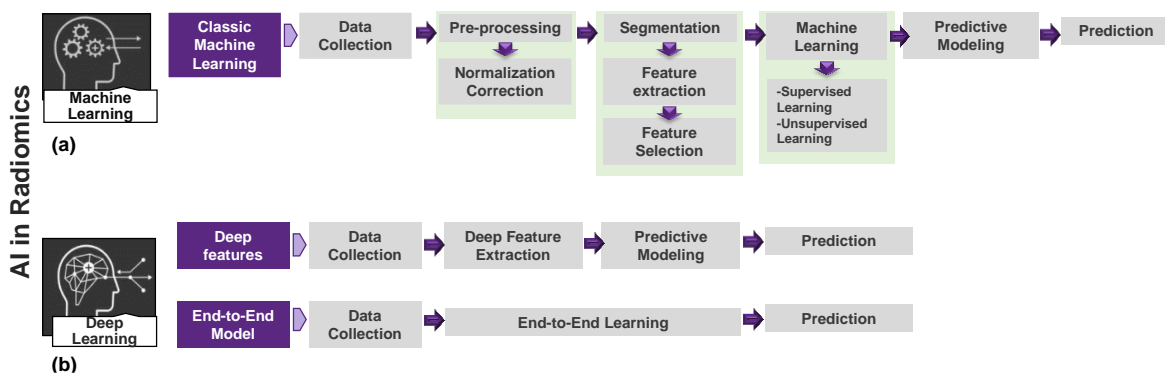

Figure 3: Typical architecture and workflow of AI systems for predictive modelling: a) classic machine learning, with the various processing steps involving handcrafted (explicit) features; b) deep-learning based techniques involving "deep features" i.e. deep image feature extraction and "end-to-end learning" (regenerated inspired by [58]);

Some explicit features are first-order or histogram-based features (such as Standardized Uptake Value(SUV) maximum, mean, peak, median, first-quartile, third-quartile, standard-deviation, skewness, kurtosis, energy and entropy) or texture features (such as gray-level co-occurrence matrix (GLCM), gray-level run length matrix (GLRLM), gray-level zone length matrix (GLZLM) and neighborhood gray-level dependence matrix (NGLDM)). Because of a potentially massive number of extracted radiomics features, radiomics model building typically involves feature selection and dimensionality reduction [49, 59]. Different approaches can be used for feature selection [60] including: (1) pre-elimination of features, such as removing features known in the literature to have poor reproducibility (e.g. overly sensitive to segmentation method), (2) using data-specific unsupervised methods (e.g. removing features that are highly correlated, or with very low dynamic range in the data), and/or (3) supervised methods for dimensionality reduction (i.e. removing features that do not add value on a separate labeled training set). Unsupervised approaches i.e. principal component analysis (PCA) or cluster analysis remove redundant features without considering the class labels [61], while supervised techniques consider class labels, and select the features based on their discriminative contribution in the specific tasks [62].

**ii) Deep radiomics**: Data-driven DL approaches are directly applied on the input image or ROI/VOI of the images, and "deep features" are learned; e.g. by utilizing convolutional neural networks (CNNs) or auto-encoder networks. DL approaches are capable of automatically identifying parts of the image most relevant to the task [63]. Auto-encoder networks, as unsupervised variant of CNNs, are able to extract compressed image content and map it onto representative features [64]. Consequently, deep features are extracted, instead of using explicit mathematical definitions of features. Despite theoretical considerations on the greater expressive power of deep features compared to explicit features, it has been shown that some explicit features are difficult to capture by CNNs of limited depth given limited training data [65, 66]. Furthermore, CNNs can be negatively biased in capturing shape information [67, 68] which can be important for a range of clinical tasks [57]. Consequently, explicit radiomic features may be complementary to deep features, and can be used as input to CNNs. Alternatively, deep features can be extracted using a CNN and then entered into a classifier [69, 70], integrated with explicit features for overall radiomics analysis. For a specific clinical task (e.g. predictive modeling), Figure 3 shows classic machine learning vs. deep learning paradigms, where in the latter, both steps of feature learning and radiomics signature derivation can be combined into one-step i.e. "end-to-end" training (Figure 3).

In the specific context of PET and PET/CT imaging, which the rest of this article focuses on, there are a number if challenges on the path to routine deployment of radiomics. There remain limited studies on multi-centric data in some cancer types such as lymphoma to investigate the importance of radiomics analysis and avoid overfitting and degraded generalizability [71]. Overall, these issues need to be carefully considered towards appropriate radiomics frameworks including for specific detection and classification tasks. The emergence of AI techniques and incorporation of radiomics signature from anatomical information of CT images can yield better models with good performance and generalization for classification, detection, and prognosis and outcome prediction. There are some challenges in PET imaging due to the inherent image quality, standardization of imaging and reconstruction that should be taken into consideration. The limited available data and limited interpretability of deep learning approaches challenge trustworthiness of AI applications in the field of medical imaging in general. In the next section, we briefly consider classification and detection techniques and provide examples in the field of oncology using this minable information of images i.e. radiomics signature that we discussed in this section.

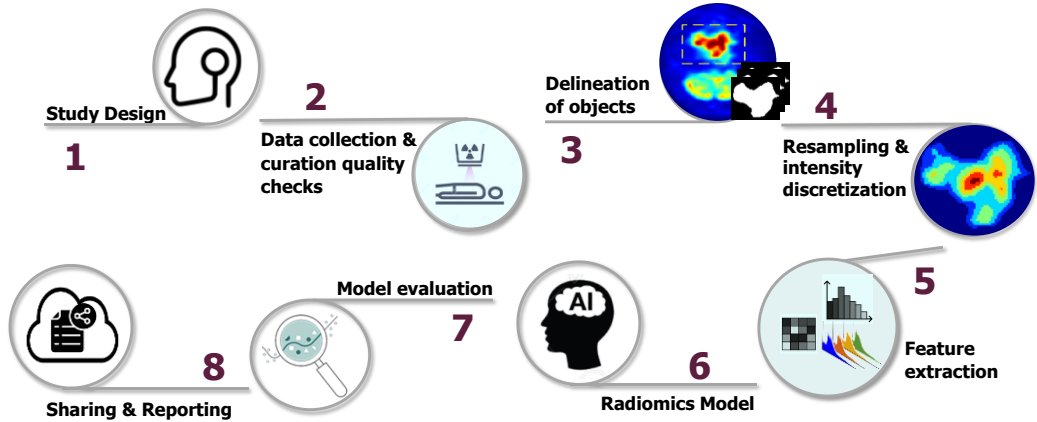

Figure 4: The general workflow for explicit radiomics analysis by AI techniques. (1) Study design step and (2) data collection sometimes can be reversed or combined. Data quality should be checked before any further analysis. (3) The features can be directly affected by ROI/VOI delineations, so this step should be done carefully by automatic or semi-automatic techniques. (4) Spatial resampling and intensity discretization (except for shape features) are performed in this step (5) Explicit (engineered) features are extracted in step 5 and (6) radiomics model is then applied. The model is evaluated and shared in the final steps (7 and 8) (inspired by [29]).

## 4. Clinical utility of AI-based Detection in PET/CT: current status

Detection of organs or lesions as regions of interests (ROIs) or volumes of interests (VOIs) is an important step towards classification of the regions [72]. Besides, most existing segmentation approaches have an embedded detection step; we discuss segmentation methods in another work [51]. End-to-end detection systems developed by AI techniques can effectively remove the need for further processing steps or prior knowledge about the detection task. Such networks can learn the most salient features based on various medical images in the training data such as PET, CT or PET/CT scans. By contrast, conventional techniques for detection have the following two stages: 1) candidate generation based on specific features of the voxels in CT or PET images; i.e. based on Hounsfield Unit (HU) or standardized uptake value (SUV); and 2) false positive reduction [73, 74]. Example attempts on the use of AI techniques for candidate generation [75] or for both steps [76] have been reported.

Convolutional neural networks (CNNs) have opened up far-reaching opportunities in medical image analysis [77-79]. The potential of using CNNs for detection is based on their ability to extract data-driven features and relationships between adjacent pixels (voxels) in 2D (3D) schemes. These hierarchical features extracted in convolutional layers of CNNs can be highly representative if sufficient training data is available[80]. There are a number of studies on detection tasks in PET or CT images; meanwhile, detection based on combined PET/CT has gained more attention recently. Table 1 lists a number of important studies involving the application of AI for detection. We review detection from PET, CT and PET/CT next.

Table 1: Selected studies on AI applications for tumor detection in CT, PET and PET/CT images

| Study | Task | Modality | Method | # of scans (n) # of cases (N) | Results |
|---|---|---|---|---|---|
| Cai et al. [81] | To harvest lesions from incompletely labeled datasets. (Composed of lung nodules, liver tumors, and enlarged lymph nodes) | CT | CNN | n=10594 N=4427 | Precision= 90% annotating 5% of the volumes ➔ harvest 9, 805 additional lesions: 47.9% recall at 90% precision, (s a boost of 11.2% in recall over the original RECIST marks) |
| Huang et al. [82] | fully-automated end-to-end system to detect & segment the lung nodule | CT | CNN Nodule detection: Faster regional-CNN (R-CNN), | n= 1018 | Detection Accuracy:91.4% and 94.6% with an average of 1 and 4 false positives (FPs) per scan |
| Zhu et al. [83] | Lung nodule detection & classification benign/ malignant) | CT | Detection: 3D Faster R-CNN Classification: gradient boosting machine | LIDC-IDRI dataset(n=1018 cases) | Detection accuracy: 84.2% Classification accuracy: 92.74 |
| Xie et al. [84] | Pulmonary nodule detection | CT | ResNet for detection DenseNet for false positive reduction, | LUNA16(n= 888) | Sensitivity: 95.3 % AFP: 1/case FROC score of 0.9226 |
| George et al. [85] | Lung nodule detection | CT | YOLO | n=880 | Sensitivity:89 % AFP: 6/case |
| Wang et al [86] | Lung nodule detection | CT | FocalMix: semi-supervised learning | n=400 | Sensitivity:90.7 % (average sensitivity) |
| Jiang et al. [87] | Lung nodule detection | CT | CNN (4- channel input: image patches enhanced by the Frangi filter) | n=1018 N= 1010 | Sensitivity:90.1/94 % AFP: 15.1/case |
| Huang et al. [88] | Lung nodule detection | CT | Local Geometric-model-based filter to generate nodule candidate &CNN to differentiate nodule/non-nodule | N=1010 | Sensitivity:90 % AFP: 5 cases |
| Ding et al. [89] | Pulmonary nodule detection | CT | Faster R-CNN-based | N= 1018 | Sensitivity:94.6 % AFP: 15/case |
| Shin et al. [80] | Thoraco-abdominal lymph node (LN) detection & interstitial lung disease (ILD) classification | CT | CifarNet, AlexNet, Overfeat, VGG16 and GoogleNet (transfer learning(from non-medical images) | n=905 N=120 | sensitivity: 85% False positive per patient: 3 |
| Vivanti et al. [90] | Automatic detection of new tumors and tumor burden evaluation in longitudinal liver CT scan | CT | CNN | No of tumors: 246 (97 new tumors, from (n=37) longitudinal liver CT studies) | TP new tumors detection rate of 86% versus 72% with stand-alone detection, Tumor burden volume overlap error of 16%. |
| Ghesu et al. [91] | Reformulating the detection problem as a behavior learning task | CT | Deep reinforcement learning in a multi-scale scheme. | n=1487 N=532 | Accuracy (mm) = 4.192* *With no failures of clinical significance |
| Xie et al. [92] | Automated pulmonary nodule detection | CT | 2D CNN (faster R-CNN) | n=1018 | (pulmonary nodule) Sensitivity=73.4% at 1/8 FPs/scan Sensitivity= 74.4% at 1/4 FPs/scan |
| Deng et al. [93] | Recognition of sFEPU and detection of lymphoma | PET | five FCNs in parallel. | n=569 | detection rate = 80.0% false positives per volume 3.2 |
| Afshari et al. [94] | Localization and detection normal active organs in 3D PET scan | PET | CNN (YOLO) | n=479 N=156 | average organ detection precision of 75-98%, recall of 94-100%, mean IOU of up to 72%. |
| Kawakami et al. [95] | Detection of abnormal and physiologic uptake region | PET | CNN (YOLO) | n= 3198 MIP N=491 | APs for physiological uptakes were: Brain: 0.993 Liver: 0.913 Bladder: 0.879 mAP = 0.831 IoU threshold value 0.5. FPS: 31.60 ± 4.66. false-positive rate=0.3704 ± 0.0213, \false-negative rate=0.1000 ± 0.0774 |
| Schwyzer et al. [96] | Detection of lung cancer in FDG-PET imaging in the setting of standard and ultralow dose PET scan | PET | deep residual neural network (RNN) | N=100 | standard dose Sensitivity=95.9% specificity =98.1% ultralow dose PET$_{3.3\%}$ Sensitivity= 91.5% specificity=94.2% |
| Blanc-Durand et al. [97] | Lesion detection in gliomas | PET | CNN(3D U-net) | N=37 | sensitivity=0.88 specificity= 0.99 |

| | | | | | positive predictive value=0.78 |
|---|---|---|---|---|---|
| Schwyzer et al. [96] | Detection of lung cancer | PET | deep residual neural network (RNN) | N=100 | Sensitivity=95.9%<br>specificity =98.1%<br>ultralow dose PET$_{3.3\%}$<br>Sensitivity=91.5%<br>specificity=94.2% |
| Bi et al. [98] | Simultaneous classification of abnormalities and normal structures in lymphoma cases | PET-CT | CNN | N=40 | F-score = 91.73% |
| Xu et al. [99] | Multiple myeloma lesion detection | PET/CT | W-net | N=12 | Sensitivity=73.5<br>Specificity= 99.59<br>Precision= 72.46 |
| Teramoto et al. [100] | Detection of solitary pulmonary nodules | PET/CT | ACM+Thresholding (initial detection)<br>CNN (feature extraction)<br>Multi-step classifier (rule based and SVM) | N=104 | Sensitivity: 90.1%<br>4.9 FPs/case |
| Kumar et al. [101] | Detection of non-small cell lung cancer (NSCLC) lesions | PET/CT | CNN | N=50 | Precision= 99.90 ± 0.11<br>Sensitivity= 97.94 ± 0.76<br>Specificity= 98.82 ± 1.33<br>Accuracy= 98.02 ± 0.71 |
| Borrelli et al. [102] | Detection of the abnormal lung lesions and calculation of the total lesion glycolysis | PET/CT | CNNs for lung lesion detection in PET and CT images and organ segmentation in CT images. | N=112 | sensitivity =90%<br>missing lesions= 1<br>positive predictive values= 88%<br>negative predictive values = 100% |
| Weisman et al. [103] | Detection of diseased lymph nodes in lymphoma patients | PET/CT | CNN (ensemble of three DeepMedic) | N=90<br>Hodgkin's (N=63)<br>diffuse large B-cell lymphoma (N=27) | TPR: 85%<br>4 FPs/patient |
| Punithavathy et al. [104] | Evaluation of the classification of lung cancer using a deep model (ResNet-18) on conventional CT and FDG PET/CT via transfer learning | PET/CT | ResNet-18 | n= 359 | Accuracy = 0.877 (CT data)<br>Accuracy = 0.817 (CT data with metadata (SUVmax and lesion size))<br>Accuracy = 0.837 (CT of PET/CT data)<br>Accuracy = 0.762 (CT of PET/CT data with metadata (SUVmax and lesion size)) |
| Zhao et al. [105] | Detection of pelvis bone and lymph node lesions | PET/CT (PSMA) | CNN (2.5D U-net) | N=193 | Bone lesion detection:<br>Precision= 99%<br>Recall=99%<br>F1 score=99%<br>Lymph node lesion detection:<br>Precision:=94%,<br>Recall=89%<br>F1 score=92% |
| Zhou et al. [106] | Neuroendocrine tumor detection | PET-CT [$^{68}$Ga] DOTATATE | ML-based algorithm (random forests regression) | N=19 | Random Forests regression machine learning model is robust to generate parametric images |

FP: False Positive
TP: True Positive
FCN: Fully Convolutional Neural Network
AFP: Average false positive
MIP: maximum intensity projection
mAP: mean average precision
IoU: Intersection over the Union
All the PET images are FDG PET except when it is indicated.

## 4.1. Detection of malignant lesions in FDG PET/CT imaging

### 4.1.1. PET-only images as input

PET imaging is used routinely and extensively for detection and characterization of lesions in cancer examinations. Detection task in PET images are mostly degraded to high-uptake region detection using thresholding approached. Use of physiological uptakes to detect abnormal uptakes has been mostly utilized to this aim.

Some authors attempted to come up with a magical SUV threshold in order to detect pathological FDG uptakes as any region with activity more than that value [107]. The threshold used can be a fixed value, e.g. SUV≥2.5 [108], or in reference to background uptake (e.g. using liver or surrounding tissue) [109]. Nonetheless, this approach is neither biologically plausible nor clinically meaningful since it ignores the normal biodistribution of radiotracer. Each radiotracer has a specific time-dependent pattern of biodistribution in healthy body. The concept of "sFEPU" ( sites of FDG excretion and physiological uptake) is misleading. Any uptake of FDG in normal tissue is "physiological uptake", whether low level/background level (as in subcutaneous fat or blood pool) or high level (as in brain). However, in the literature the term sFEPU only refers to the physiological uptakes that are above certain level of activity (i.e. above background conceptually). The term high normal activity (HiNA) is a better term (ND=LoNA+HiNA)

The attempts to find a spatially-invariant/global threshold to detect "sFEPU" (a.k.a. HiNA) is destined to fail based on biological concepts; the poor performance of these methods has been demonstrated empirically multiple times. Moving beyond spatially-invariant thresholding, Bi et al. [98] proposed a multi-scale superpixel encoding approach that first groups the fragments that are obtained based on a multi-scale superpixel-based encoding method (MSE) and a class-driven feature selection and classification model for sFEPU classification in of whole-body lymphoma PET-CT scans. The features are then extracted using domain-transferred CNNs to classify the regions into sFEPU and anomalies. Deng et al. [93] proposed a network composed of two parts for the detection of lymphomas and sFEPU. Their model consists of five fully connected convolutional neural networks (FCNs) in parallel. The input image is resized into five different spatial scales (100%, 95%, 90%, 85%, 80% of PET image dimensions), and each is entered as an input to the corresponding FCN model. In the second step, the detected sFEPUs in different scales are integrated to determine the exact sFEPU boundary.

Afshari et al. [94] and Kawakami et al. [110] applied the deep convolutional neural network (CNN), You Only Look (YOLO) [111], to detect HNA on 2D PET images. YOLO detected multiple organs in 2D slices of the maximum intensity projection (MIP) images. Afshari et al. [94] then aggregated the detection results on 2D images to produce a 3D probability map by. The size and metabolic levels of lesions impacted the performance of tumor detection techniques.

There are some attempts to augment the training data for detection tasks by generating synthetic PET. Generative adversarial networks (GANs) can utilize the existing CT data to synthesize PET images with high uptake region and constrain the appearance of the generated PET images [112]. This data augmentation method may help the model with false positive; in other words, applying thresholding to the synthesized PET image to extract high-response regions can reduce the false positive rate by comparing the detection and thresholding masks [113, 114].

### 4.1.2. CT-only images as input

Most existing studies for detection in CT imaging applied DL (more specifically CNNs) in one of the following ways: (i) training CNNs from scratch or (ii) applying CNNs pre-trained on natural or medical images via transfer learning. 3D networks for supervised learning require more training data, which is commonly a challenge in the field of medical imaging due to the relatively limited availability of labeled data. This is effectively a "curse of dimensionality" problem, resulting in overfitting of sparse data in 3D approaches [115]. Overall, detection based on 3D networks is non-trivial, and 2D or 2.5D networks (i.e. using the information from only few adjacent slices) have been a commonly explored alternative given sparsity of data. Unsupervised learning is also another noteworthy approach for detection; e.g. Afifi et al. [116] utilized graph cuts, and iteratively estimated shape and intensity constrains for liver lesion detection in CT images.

Shin et al. [80] exploited a combination of training and transfer learning (from non-medical images) to detect thoraco-abdominal lymph nodes and for interstitial lung disease classification in CT images. They used various CNN architectures (CifarNet, AlexNet, Overfeat, VGG16 and GoogleNet) with different number of layers and variation of parameters.. They specifically explorer the capability of three important approaches that employ CNNs to medical image classification for detection and classification: (i) training the CNN from scratch, (ii) applying off-the-shelf pre-trained CNN features, (iii) unsupervised CNN pre-training with supervised fine-tuning, (iv) transfer learning, i.e., fine-tuning CNN models pre-trained from natural image dataset to detection task. They results showed that CNN model can be used for high performance CADe systems.

A 3D approach for localization of anatomical structures was proposed by de Vos et al. [117] based on 2D detection scheme from orthogonal planes using independent CNNs. Their proposed method worked well for localization of structures with clear boundaries. Vivanti et al. [90] proposed a detection method for liver tumors burden quantification in longitudinal liver CT images. Their proposed method detects new tumors in the follow-up scans and quantifies the tumor burden change using a CNN as a prior model without the need of large labeled training data. Ghesu et al.[91] tried to solve the detection problem as a behavior learning task for an AI system. They defined a unified behavioral framework based on deep reinforcement learning in a multi-scale scheme. The network was trained to detect the anatomical object using an optimal navigation path to the target object in the 3D volume of data.

Xie et al. [92] suggested a detection framework based on 2D CNN for pulmonary nodule in CT images. The Faster R-CNN was adjusted with two region proposal networks and a deconvolutional layer were employed to detect the nodule candidates. Three models were trained for three kinds of slices to be fused in later stages. A boosting architecture based on 2D CNN were used to reduce the false positive (FP). The misclassified samples are then used to re-train a model to boost the sensitivity. The results of these networks were finally utilized to vote out the result. Their proposed detection framework is shown in Figure 5.

Some of the AI models for cancer diagnosis applications can be interpretable by determination of the most influential regions in the input images that determine the output of the model i.e. saliency map [118]. While most of the existing detection and segmentation networks, such as U-Net and Faster R-CNN need annotated data to be trained and such supervised models are prone to be biased to the distribution of the training data if they are not diverse enough. Weakly supervised techniques, have gained much attention to overcome these limitations using only image-level annotations to locate ROIs in an image based on saliency map that encodes the location of ROIs [35, 36]. For instance weakly supervised techniques have been applied for cancer detection in lung CT images [119, 120].

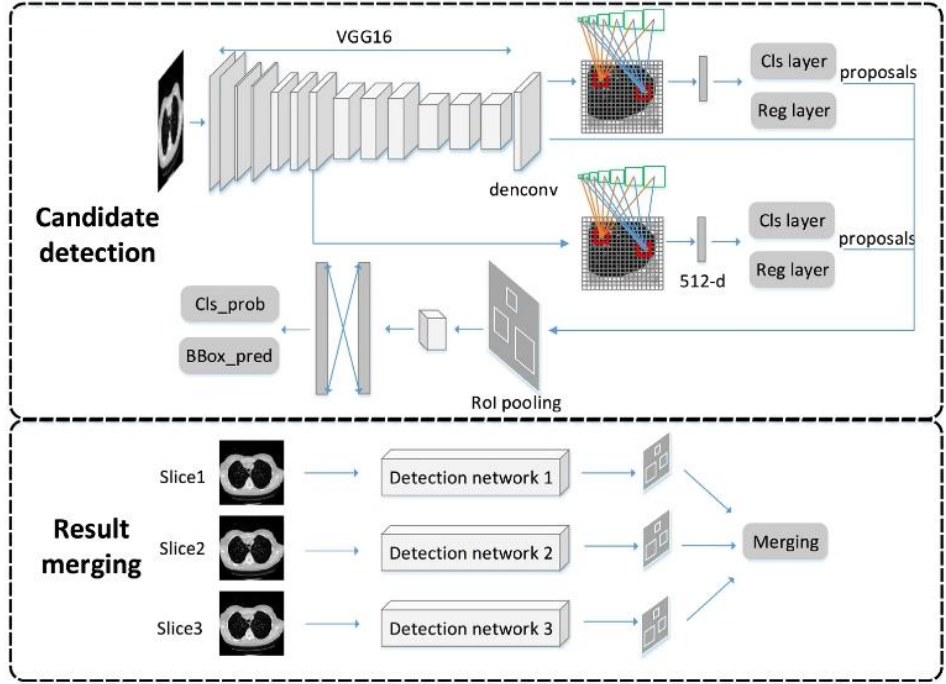

Figure 5: The framework of the nodule candidate detection. The basic feature extraction network is VGG16 and a deconvolution layer is used to enlarge the feature map. Meanwhile two region proposal networks with designed seven ancjors are applied to obtain the proposals. Finally ROI classified is employed to get the candidates (adapted from [92]).

### 4.1.3. Combined PET and CT images as input

PET/CT imaging as a powerful tool for the accurate diagnostic of oncological patients are widely used for prognostication and therapy response assessment [121]. Some existing studies used the spatial information of CT and metabolic information of PET images to develop an automated detection framework. For example, to detect small lung nodules (<4 mm) in clinical workflow, the wait-and-see approach is suggested. Considering the risk of delayed treatment and biopsy, FDG PET/CT is recommended by the society of nuclear medicine recommends for diagnostic [122]. Tumor detection based on PET/CT images are performed generally by extracting the high SUV region and involving its information in the detection scheme based on CT images. Xu et al. [99] utilized two CNNs for multiple myeloma bone lesion detection and segmentation from $^{68}$Ga-Pentixafor PET/CT. They used two-cascaded V-net i.e. W-net; the first V-net model is fed by CT images and trained to learn the bone anatomical information. The output of the first V-net is a binary mask for the skeleton anatomical map. Both PET and CT data are fed into the second V-Net to detect the lesion. They showed that a W-net architecture, which combined the extracted features from the two-cascaded V-nets on PET and CT images, outperformed the V-nets on each modality for both lesion detection and segmentation tasks. Moreover, the CNNs outperformed conventional machine learning (ML) techniques such as random forest (RF), k-nearest neighbors (kNN) and support vector machine (SVM).

Teramoto et al. [100] first located the candidate lung nodules separately by active contour model (ACM) and thresholding in CT and PET images. The results of these two streams are then combined using logical OR function. An ensemble of i) a multi-step classifier utilizing two feature-extractor for CT and PET and ii) a CNN followed by a two-step classifier and SVM were then applied. Their results revealed that the feature extracted by CNN (they used three convolution layers, three pooling layers, and two fully connected layers) help the FP reduction. Kumar et al. [101] utilized a CNN to facilitate the spatially varying fusion of complementary information of PET and CT images of non-small cell lung cancer (NSCLC) using an encoder for each modality, a co-learning and a reconstruction components. The results showed the better performance of their proposed method compared to other CNN-based methods for bi-modal detection of lungs, mediastinum and tumors. Blanc-Durand et al. [97], using amino-acids PET with $^{18}$F-fluoro-ethyl-tyrosine ($^{18}$F-FET), showing that a 3D U-net is able to successfully detect brain gliomas. Weisman et al. [73] utilized the multiresolution pathway CNN, DeepMedic for lesion detection in lymphoma patients. An ensemble of fivefold cross-validation model were applied and performance was assessed using the true-positive rate (TPR) and number of FP.

More recently, low-dose PET/CT has shown increasing potential for first-line screening. In a preliminary study, Schwyzer et al. [96] showed that using a transfer learning approach, a pre-trained deep residual neural network (RNN) for binary classification can be used for fully automated lung cancer detection utilizing FDG-PET scans with very low effective radiation doses (ultralow doses: thirtyfold (PET3.3% )) [96]. They also showed their network had the potential to aid in detection of small FDG-avid pulmonary nodules (≤ 2 cm) [121]. The results revealed that DL models could help in detection of small $^{18}$F-FDG-avid pulmonary nodules in PET/CT scans. Interestingly, they noted their DL method to perform significantly better on images with block sequential regularized expectation maximization (BSREM) reconstruction as compared to ordered subset expectation maximization (OSEM) reconstruction [121].

### 4.1.4. Detection of malignant lesions in non-FDG PET/CT

Edenbrandt et al. [123] developed an AI tool using CNN for detection and quantification of primary prostate tumors, bone metastases and lymph nodes in PSMA PET/CT. They showed that utilizing their AI tool decreased the influence of inter-reader variability. Borrelli et al. [102] employed two CNNs for lung lesion detection in 18F-choline PET/CT images and organ segmentation in CT images. They used the segmented organs as an auxiliary input to the detection network (Figure 6). They reported the sensitivity of 90% for their lesion detection approach. AI techniques have recently shown a very good potential to help detect metastases in $^{18}$F-FACBC (fluciclovine), 68Ga-PSMA-11 and $^{18}$F-choline PET/CT scans of prostate cancer patients [105, 124-126]. We will discuss more details of these studies in classification section.

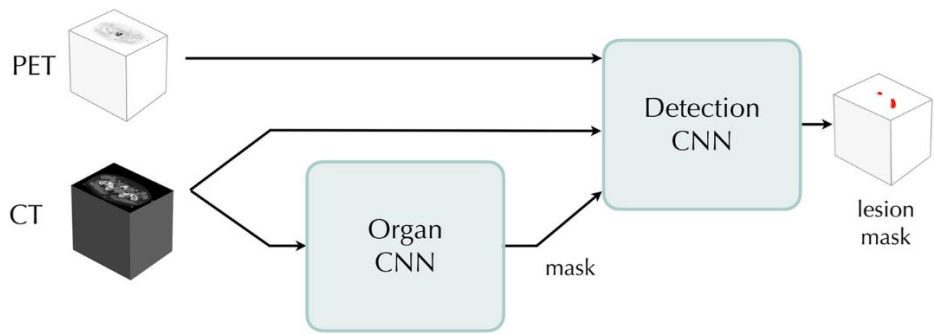

Figure 6: The AI technique proposed by Borrelli et al. [102] based on two CNNs. The "Organ CNN" segments the lungs, vertebral bones, liver and heart. The mask of the organs thus generated are used as auxiliary input to the "Detection CNN", in addition to the PET and CT images. The output consists of lesion probabilities that can be thresholded to produce a lesion mask (adapted from [102]).

# 5. Clinical utility of AI-based classification in PET/CT: current status

Classification is considered as the most popular area in which CNNs have been used; e.g. AlexNet, ResNet, DenseNet, VGG network, and others [79]. Similarly, in medical imaging, AI techniques have been widely employed for extraction of feature towards: (i) classification of suspicious lesions and tumor subtypes, as well as (ii) prediction/prognostication tasks, stratifying/classifying patients into risk groups [127, 128]. We review both frontiers next.

## 5.1. Lesion classification in oncological PET imaging

Lesion classification can be very challenging and has been conventionally attempted via thresholding. AI techniques seek to enhance such tasks. For instance, abnormal foci in $^{18}$F-FACBC [fluciclovine] PET/CT images and the presence of recurrent or metastatic prostate cancer can be classified automatically using AI techniques [124]. Table 2 lists a number of example studies utilizing AI for classification. We review application for both FDG and beyond.

Table 2: Selected studies on AI applications for classification in oncological PET and PET/CT imaging

| Study | Task | Modality | Method | # of scans (n) # of cases (N) | Results |
|---|---|---|---|---|---|
| Lee et al. [124] | Discriminating normal and abnormal scans | PET $^{18}$F-FACBC | 2D-CNN (ResNet-50) and a 3D-CNN (ResNet-14) | n=251 N=18 | 2D: Sensitivity=85.7% specificity=71.4% AUC=0.750 3D: Sensitivity=71.4% specificity=71.4%, AUC=0.699 |
| Leung et al. [129] | Classification of PCa lesions | PET (PSMA) | 3D CNN | N=267 | Accuracy= 67.3% |
| Teramoto et al. [130] | Classification of pulmonary nodules | PET/CT | a rule-based classifier and three support vector machines | N=100 | Sensitivity: 83% FP/case: 5 |
| Sibille et al. [131] | Classify uptake patterns in patients with lung cancer & lymphoma | PET, CT & PET MIPs | CNN | N=629 | Accuracy: CT: 78% PET: 97% PET/CT: 98% PET MIP: 98% |
| Kawauchi et al. [132] | Classify whole-body FDG PET as 1) benign, 2) malignant or 3) equivocal. | PET/CT | CNN (ResNet) | N=3485 | Accuracy: Benign=99.4% Malignant= 99.4% Equivocal = 87.5% probability of correct prediction: 97.3% (head-and-neck), 96.6% (chest), 92.8% (abdomen) and 99.6% (pelvic region) |
| Moitra et al. [133] | Staging and grading of NSCLC | PET/CT | CNN | N=211 | Accuracy: T stage: 96% (2%) N stage: 94% (2%) M stage: 99% (2%) Grading: 95% (3%) |
| Perk et al. [126] | Bone lesion classification | PET/CT | ML | N=37 | sensitivity = 88% specificity = 89% |
| Acar et al. [134] | Differentiating metastatic and completely responded sclerotic bone lesion in prostate cancer | PET/CT PSMA | Decision tree, Discriminant analysis, SVM, kNN, ensemble classifier | N=75 | area under the curve (kNN): 0.76 |
| FP: False Positive SVM: Support Vector Machine MIP: Maximum intensity projection kNN: k-nearest neighbor TLG: total lesion glycoses | | | | | |

### 5.1.1. FDG PET imaging

ML techniques such as random forest can be used to classify the extracted ROIs into normal and abnormal [135]. Teramoto and Fujita et al. [130] developed an algorithm to characterize lung nodules by combining CT analysis and binarized PET images based on SUV thresholding. False positives (FPs) among the leading candidates were eliminated using a rule-based classifier and three support vector machines (SVMs). Traditional ML techniques generally use domain-specific explicit features that require domain expertise while DL-based approaches, such as CNNs, can learn abstract features directly from the training data. Some studies evaluated different CNN configurations to classify the uptake patterns to suspicious and nonsuspicious for cancer from whole-body 18F-FDG PET/CT images. Kawauchi et al. [132] utilized a RES-net for whole-body FDG PET classification to benign, malignant or equivocal.

The proper diagnosis in lung cancer is based on tumor staging and grading. For solid tumors, medical imaging is often used to determine the T (tumor), N (node) and M (metastasis) classification defining the stage of the disease. ML algorithms on PET/CT have been developed for lung cancers, notably T staging, utilizing different methods including random forests, SVM and CNN. Moitra et al. [133] proposed a 1D CNN for lung cancer staging on a public (PET/CT) dataset in the cancer imaging archive (TCIA). The application of 1D CNN for staging and grading of lung tumors is not frequent. Their 1D approach was applied on the extracted features from the segmented regions and the histopathological grade information. This 1D CNN approach could offer a lightweight solution that acts as a decision support system for oncologists and radiologists. Unlike the 2D and 3D networks that used the spatial attributes directly from the 2-D or 3-D images and generate huge resource consuming architecture. Their proposed approach used the spatial information in a rank 2 tensor dataset (a CSV file) along with clinical staging and grading data for classification. Capobianco et al. [136] used a prototype software (PET Assisted Reporting System [PARS]) to automatically detect and segment the high-uptake regions. The resulting ROIs were further processed by a CNN model to be classified as nonsuspicious or suspicious uptake. The CNN model was pre-trained on an independent cohort consists of patients with lung cancer and different subtypes of lymphoma [131, 137]. Sibille et al. [131] also proposed the use of PARS algorithm to evaluate each extracted ROI independently and determine whether an uptake is suspicious or benign. They used a combination of PET, CT, and PET MIPs, and atlas positions to classify uptakes in patients with lung cancer and lymphoma (Figure 7). An example of PARS system applied on lymphoma data is presented in Figure 8.

Du et al. [138] investigated a variety of ML techniques for radiomics-based differentiation of local recurrence versus inflammation in post-treatment nasopharyngeal carcinoma post-therapy (NPC) PET/CT Images to be applicable for clinical decision making. They showed that the cross-combination fisher score (FSCR) + k-nearest neighborhood (kNN), FSCR + support vector machines with radial basis function kernel (RBF-SVM), FSCR + random forest (RF), and minimum redundancy maximum relevance (MRMR) + RBF-SVM outperformed the other ML techniques and their combination. Du et al. [139] also considered the discrimination power of the individualized radiomics nomogram in both the training and independent validation cohorts. They showed that considering PET, CT, and PET/CT diagnostic signatures, that combination of metabolic characteristics of PET and anatomical information of CT improved diagnostic performance (Figure 9). On the other hand as Figure 9 shows a model that uses radiomics and clinical semantic features in a radiomics nomogram outperformed the model relied on PET only or CT only information.

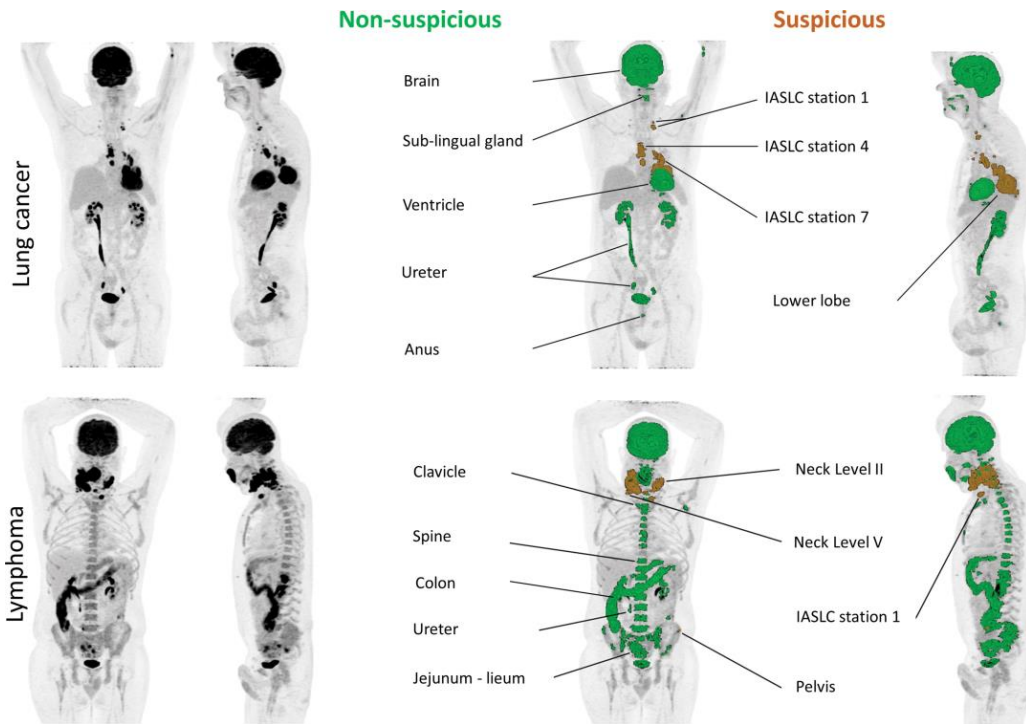

Figure 7: CCN applied on the maximum intensity projection fluorine $^{18}$F-FDG PET/CT images of two patients. Patients with lung cancer and lymphoma with areas of uptake automatically color-coded by their classification and localization are shown. (IASLC=International Association for the Study of Lung Cancer)(adapted from [131])

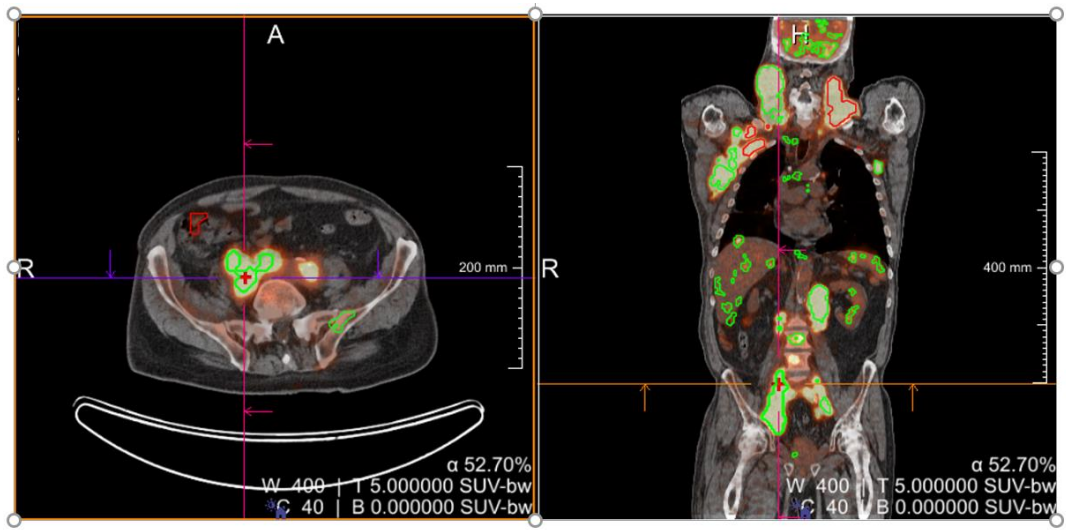

Figure 8: a patient with diffuse large B-cell lymphoma (DLBCL). Green foci are classified as benign by the prototype software (PET Assisted Reporting System [PARS]) and red foci are classified as pathological. Here, the algorithm has largely underestimated the pathology with an estimated volume in automated segmentation of 242 ml (SA) against 1270 ml in manual segmentation.

### 5.1.2. Radiotracers beyond FDG

Although $^{18}$F-FDG is the most widely used radiotracer in oncology and nuclear medicine, it is not as suitable for certain cancers, although the data access for tracers much less common than FDG, is limited. As an example, prostate adenocarcinomas, the most common cancer in men, can be imaged using other tracers such as radiolabeled choline, $^{18}$F-Fluciclovine and $^{18}$F-Piflufolastat (targeting prostate-specific membrane antigen (PSMA)) or similar radiotracers such as $^{68}$Ga-PSMA [140]. Promising results of VISION trial [141] closed the evidence gap in prostate cancer radiopharmaceutical therapy with $^{177}$Lu-PSMA. At the dawn of this new era, development of new imaging biomarkers for retreatment patient selection is in the horizon.

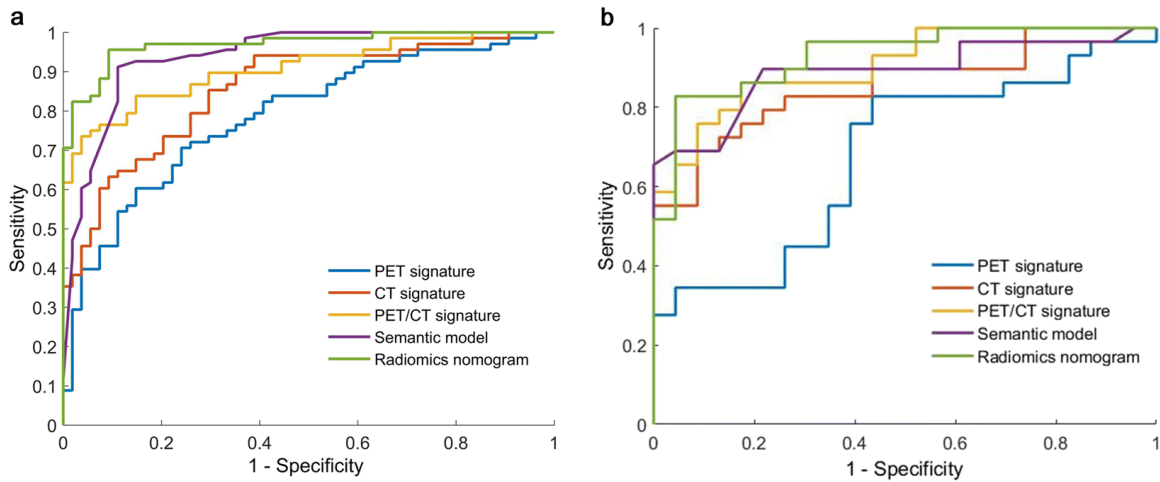

Figure 9: The receiver operating characteristic (ROC) curves of PET, CT, PET/CT signatures, semantic model, and radiomics nomogram for (**a**) training and (**b**) validation cohorts. Compared with CT signature, the PET/CT signature significantly improved AUC for training cohort. Combining the PET/CT features with the semantic features in a radiomics nomogram significantly improves the diagnostic performance with respect to the semantic model alone or PET/CT signature alone in training cohort, whereas did not significantly improve the performance in validation cohort. The DeLong test method was used for the statistical comparison of ROC curves(Figure is adapted from [139]).

Theranostic radio-isotope pairs for diagnosis ($^{18}$F, $^{68}$Ga) vs. therapy ($^{177}$Lu, $^{225}$Ac) provide us with possibility of having diagnostic and therapeutic radiopharmaceuticals with relatively similar biodistributions and kinetics. Several studies have already investigated the contribution of AI techniques to the analysis of PSMA PET/CT scans (see [142])

Zhao et al. [105] in a pilot study focused on the pelvis, developed a framework to characterize the local and secondary prostate tumors in the lymph nodes and bones and to facilitate the optimization of PSMA directed radionuclide therapy by using a 2.5D U-net architecture. . Polymeri et al. [125] used a CNN to segment prostate cancer and prostate volume based on $^{18}$F-choline and found the derived features from PET/CT images were significantly associated with overall survival. They found that the other common clinical features including age, prostate-specific antigen, and Gleason score did not affect the overall survival. Leung et al.[143] developed a 3D CNN to automatically classify the PCa lesions in $^{18}$F-DCFPyL PSMA PET images by PSMA-RADS score prediction for lesion classification. The 3D PET images were cropped to yield varied size cubic volumes-of-interest (VOIs) around the center of each lesion as the input to the 3D CNN along with the manual segmentation. This classification approach showed promising results.

Perk et al. [126] developed an automated bone lesion classification by using ML techniques to classify lesions in $^{18}$F-NaF PET/CT images, as bone metastases can confounded by tracer uptake in benign diseases, such as osteoarthritis. Acar et al. [106] used ML techniques including decision tree, discriminant analysis, support vector machine (SVM), kNN, and ensemble classifier on the textural features to distinguish the lesions imaged via $^{68}$Ga-PSMA PET/CT as metastatic and completely responded. The study population contains patients with known bone metastasis and previously treated patients and the weighted kNN outperformed the other classifiers. As it is considered in the above-mentioned studies, it has been shown that using the hybrid information of CT and PET images statistically significantly improved the lesion classification accuracy.

Lee et al. [124] utilized a 2D-CNN (ResNet-50) and a 3D-CNN (ResNet-14) to discriminate normal and abnormal PET scans based on the presence of tumor recurrence and/or metastases in patients with prostate cancer and biochemical recurrence in $^{18}$F-FACBC (fluciclovine) images. They reported that a 2D slice-based approach depicted better results than 2D or 3D case-based approaches. [$^{68}$Ga]DOTATATE PET/CT scans have shown significant value for imaging neuroendocrine tumors (NETs), improving detection, and characterization, grading, staging, and predicting/monitoring the responses to treatments, although the AI techniques for detection have not been explored.

## 5.2. Prediction/prognosis in oncological PET imaging

Prognostic tasks provide information about patient outcome, regardless of therapy, whereas predictive tasks consider the effect of therapeutic interventions [144]. The prognostic value of pretreatment $^{18}$F-FDG-PET/CT images for treatment planning and decision support has been extensively considered [145, 146]. Outcome prediction includes extraction of radiomics feature from PET and CT images, combined with any valuable clinical indicators, to predict survival, e.g. recurrence-free, metastasis-free, progression-free and/or overall survival [146-148].

Cottereau et al. [149] showed that "lesion dissemination" is a strong prognosticator of progression free survival (PFS) and overall survival (OS) in diffuse large B-cell lymphoma (DLBCL) cases. Their work emphasizes that outcome prediction can consist of finding links amongst tumors. They showed that the largest distance between two lesions (Dmax) that indicates the spread of the disease is independent and complementary to metabolic tumor volume (MTV)) for outcome prediction in DLBCL patients. Desseroit et al. [150] considered features in four categories: (1) clinical variables, (2) volume and standard metrics, (3) PET heterogeneity and (4) CT heterogeneity (Figure 10). They showed that tumor heterogeneity quantified with textural features on the CT and PET components of $^{18}$F-FDG PET/CT images can provide complementary prognostic value in Non-small cell lung cancer (NSCLC). They designed a four-variable nomogram that outperformed the standard clinical staging.

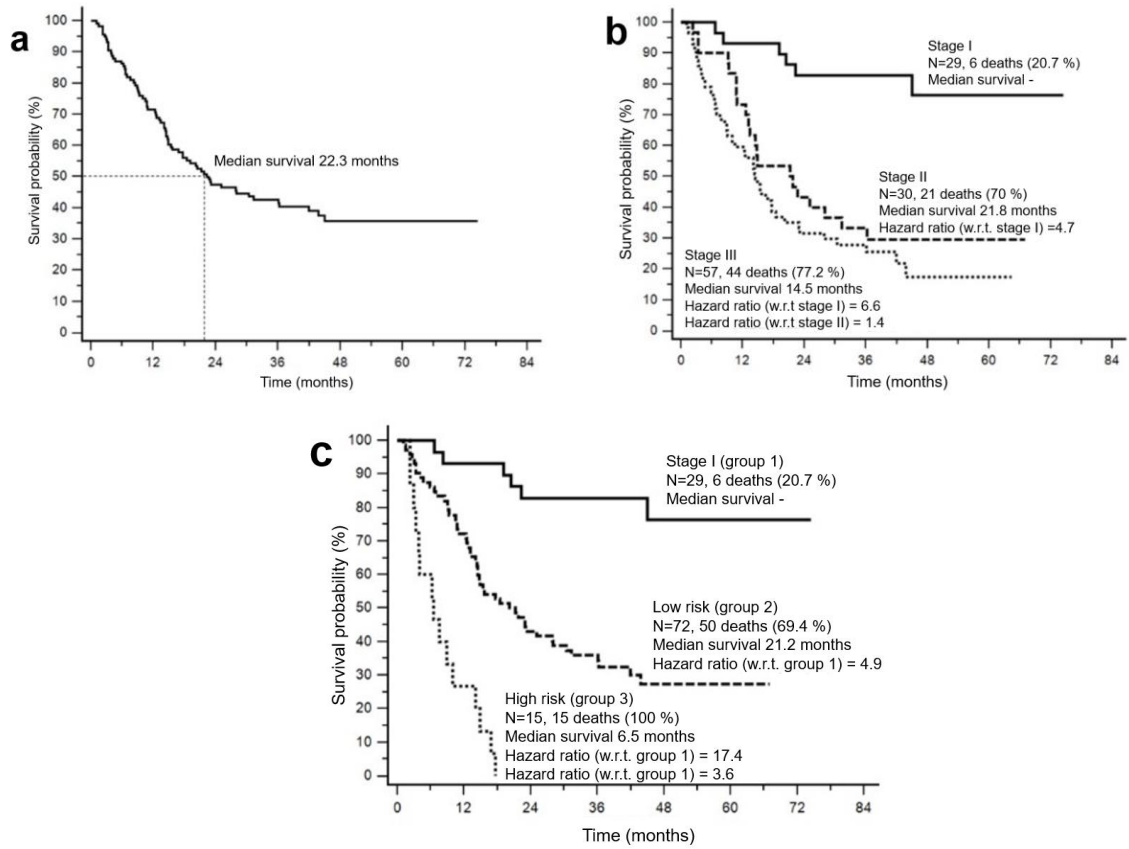

Figure 10: Kaplan-Meier curves for (**a**) the cohort and (**b**) in relation to stage only, and (**c**) the proposed nomogram (w.r.t. in the figures means with respect to). (**a**) The goal of this study is to build a nomogram combining the best features of each category in order to improve the stratification provided by stage alone. (**b**) Patients with stage III disease had worse survival than those with stage II disease. (**c**) The best stratification was obtained by including all four parameters in the model (Figure is adapted from [150]).

The hope is that AI methods can enable robust risk-stratification, thus enabling improved staging of patients. In this subsection, we consider the important applications of AI in prognosis of cancer recurrence rates, outcome prediction for response assessment and clinical decision-making. Effective prognostic signatures include both clinical and imaging features by integrating quantitative voxel-wise features and valuable clinical information. The increasing availability of high-throughput omics datasets from large patient cohorts has facilitated the development of AI techniques classify patients according to survival or disease recurrence. Specifically, for stratification of cancer patients, their death or other survival information (e.g. progression-free survival) are considered as endpoints.

Existing **predictive modeling frameworks** may perform: (1) statistical survival analysis on explicit features; (2) ML based techniques applied to explicit and/or deep features; and (3) DL (CNN) models for outcome prediction. In addition to deep feature extraction, neural networks can model nonlinear survival data by classifications and estimation of the risk of failure as the nonlinear extension of Cox proportional hazards [151]. Recent studies have shown the superiority of ML techniques over statistical analyses in terms of discriminant power utilizing a feature selection scheme based on complementary features [44, 152]. Prognostication (prognostic stratification) has been primarily determined based on semantic tumor staging, leading to relatively coarse and discrete stratification [153]. As such, direct outcome prediction without the intermediate stages of image/tumor classification has gained increasing attention. CNNs are able to learn the prognostically relevant features of the images directly by bypassing the intermediate steps for identification of predictive biomarkers. Explicit feature based ML techniques for outcome prediction usually suffer from the feature selection step that is often unstable given the high-dimensional features (curse of dimensionality), though there is ongoing research on using of dimensionality reduction algorithms (feature selection or extraction algorithms) [154]. It has been shown that CNNs have enhanced predictive power compared to explicit features [155, 156].

Ypsilantis et al. [157] considered both of the above-mentioned framework i.e. ML and CNN approaches to predict outcome in patients with esophageal cancer. The first approach included feature extraction and using ML classifiers such as linear regression (LR), gradient boosting, RFs and SVMs. The second approach applied CNNs on image data and corresponding segmented region that had higher average accuracy. Pereira et al. [158] suggested a pipeline for response to treatment evaluation and stratification into a 5-classe Deauville scale using different CNN architectures on patient data with Hodgkin's lymphoma. An artificial intelligence method suggested by Capobianco et al. [159] can generate a TMTV value prognostic of outcome in a large series of patients with non-Hodgkin's lymphoma (i.e. DLBCL). On a database of 97 patients, Bizzego et al. [69] showed that the analysis of PET tumor images with a 3D CNN produced very promising results to predict treatment response in esophageal cancer, and outperformed 2D CNN architectures, as well as explicit feature extraction with RF classifiers. The hybrid– explicit and deep feature approach was the most accurate. The same authors proposed to carry out internal transfer-learning based on the prediction of the grade of the disease to improve the prediction of their prognostic model and have shown a better classification accuracy using this transfer learning approach. On the other hand, Wang et al. [160] showed that CNN results were not superior compared to explicit radiomics for mediastinal lymph nodes classification in non-small lung cancer data. They concluded that explicit features are sometimes preferred, since they are user-friendly and can be applied with less amount of data and are less affected by feature selection bias.

# 6. Future Directions & Open Questions

## 6.1. Temporal changes in Radiomics Features: Dynamic-Radiomics and Delta-Radiomics

Radiomics signatures of medical images are usually derived from static images. It has been shown that temporal changes for example in tracer uptake of PET scans can also reveal new aspects of tumor biology [161-164]. Analysis of features extracted from temporal analysis of dynamic PET scans can be referred to as "dynamic-radiomics" (micro-scale temporal changes). By contrast, "delta-radiomics" refer to analysis of features derived from comparison of a study with prior images (to capture macro-scale temporal changes). Clinical experience highlights the importance of temporal changes as one of the most important characteristics of the lesion [165]. Fave et al. [163] showed the utility of CT delta-radiomics for prediction of patient outcome in lung cancer [163]. Later, numerous groups expanded the use of CT-based delta-radiomics to other cancers such as gastric cancer [166], and also detection of treatment side-effects such as radiation induced xerostomia [167, 168] and pneumonitis [167]. MR-based delta-radiomics is also widely utilized in various cancers, including prostate cancer [169], sarcoma [170], colorectal cancer [171], and rectal cancer [171]. It has been suggested that delta-radiomics may increase multicentric reproducibility [172]. As such, delta radiomics frameworks are being developed (e.g. utilizing pre-, intra-, and/or post-therapy scans in combination. Such a framework also has potentially significant value when applying to theranostics paradigms in the context of radiopharmaceutical therapies [142].

## 6.2. Efforts to Translate AI Technique to Routine Clinical Usage

Translating AI techniques into routinely employed clinical workflows requires collaboration between AI researchers, radiologists and oncologists, and predefined frameworks for evaluating these techniques to be integrated into clinical applications and accepted by the oncology community [128]. The decisions made by nuclear medicine physicians and radiologists are not based on 2D slices of PET or CT scans. Some of the existing studies suggested the AI techniques applied on the limited 2D slices for detection task that can be very prone to errors and considering the advanced 3D AI techniques, it should be re-considered in future studies in the field of PET oncology.

Explainable AI techniques such based on attention or saliency maps are demanded especially for clinical decision support systems [31, 173, 174]. The explanations is needed to: i) verify the model's output to the physicians, ii) increase the trustworthiness of AI techniques and iii) identify the potential biases [175, 176]. Consequently, explainable AI techniques facilitate the communication between the "black-box" DL models and physicians.

Reproducible measurements made by AI methods for lesion evaluation in oncology reports have been shown to be helpful for reducing the reporting errors and increasing the workflow efficiency [177]. Considering the lower FPs and high sensitivity of DL based methods compared to traditional ML techniques for detection, DL based methods can be clinically feasible if they showed similar performance in the extensive clinical trials.

AI frameworks have the potential to be used as a "second reader" for radiologists to improve diagnostic accuracy and efficiency [127, 128]. It is worth noting that AI-based solutions (considering the superior performance of DL approaches) can be used as the decision support tools while they have been mostly considered as the replacement of clinicians in clinical practice that is far from reality [178]. In fact, AI techniques try to mimic and augment the clinicians' work instead of replacing them [179]. Freeman et al. [180] showed that the promising results of AI techniques in smaller studies of breast cancer screening programmsare not replicated in larger studies.

Most studies to date on radiomics signature have involved limited number of cases (<100 prone to overfitting), retrospective (selection bias) and monocentric (less likely to be generalized); the last limitation refers to the fact that most of the developed models were never evaluated on external datasets [181]. Furthermore, it has been demonstrated that despite the growing number of developed frameworks for disease severity classification (estimation), the existing models have been reported to be largely dependent on the demographics and clinical history of the patient [182].

Challenges with existing AI techniques include the limited interpretability of the resulting networks and the high dependency on large datasets in supervised techniques. Weisman et al. [73] investigated the impact of the number of training subjects on detection performance; their results showed that it is not always the case that using a much larger dataset can improve performance of supervised AI techniques and that the performance of the DL model is limited by the consistency of the labels (Figure 11). In the other words, sample size, data quality and diversity are all important [183]. New AI techniques that need the lower amount of annotated are highly demanded e.g. semi-supervised and weakly supervised techniques that use sparse and weak annotations showed acceptable performances. Utilizing the annotations made by multiple experts for every scan was proposed to produce limited data with high quality annotations that are not achieved usually [31]. The platforms for "crowd-sourced" image labeling can help in this regard [184]. The number of cases needed for developing a radiomics signature model is highly dependent on the complexity of the problem [185]. In addition to use of unsupervised and weakly supervised techniques, using DL models pretrained on different pathologies can also be helpful [186]. Capobianco et al. [187] showed that using transfer learning the CNN network trained for tumor detection in $^{18}$F-FDG PET/CT can be applied for tumor assessment in $^{68}$Ga-PSMA-11 PET/CT images.

The nuclear medicine community is now aware of the impact of image acquisition and reconstruction parameters on handcrafted radiomic features (and on the performance of models combining them) but this issue has been less discussed and reported in deep radiomics while there is no guarantee that the effect is not as strong and more difficult to detect. A number of problems need to be tackled for clinical application of DL-based techniques. For instance, the performance of the model is highly affected by the curse of hyper-parameters [188] (a range of parameters to optimize such as learning rate, number of epochs, activation functions, etc. that usually are chosen arbitrarily). The lack of explainable model for an AI-based decision support system in clinical application is another issue that prevents rapid translation of AI techniques into clinical workflow [189].

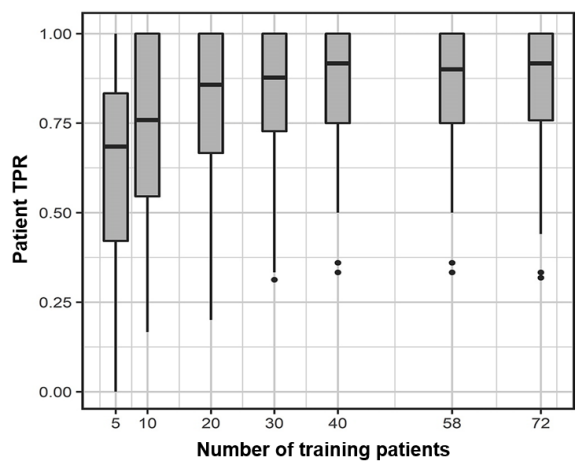

Figure 11: Considering the impact of the number of training patients on detection performance. The detection performance was not improved when training was done on more than 40 subjects. No significant differences (p>0.05) were observed in the performance of the model trained on N=40 subjects compared to the model trained with 58 subjects. Likewise, model performance did not significantly improve when training with 72 subjects (using the validation cohort as additional training subjects). Boxplots show median and interquartile range of patient-level true-positive rate (TPR) across all 90 patients as a function of training patient number, operating at an average of four false-positive findings per patient. Dots represent outliers, which are determined by points lying further than 1.5 times the interquartile range from the edge of the box (adapted from [73]).

### 6.3. Application of natural language processing for labeling

The extracted information from clinical reports can be used to develop AI techniques that can help reporting in clinics. Although there are guidelines for diagnostic imaging reports and label extraction, the majority of these reports are composed of free-text [190]. Rule-based [191] or recurrent neural networks [192] for natural language processing (NLP) are used for information extraction from radiology reports. In the field of NLP, non-experts mostly do the manual labeling; this approach is not applicable in the field of medical image segmentation especially for advanced modalities i.e. CT and PET. Recently, reporting that contains hypertexts [193], labeling before the radiologist i.e. prospective labeling [194] and interactive reporting [195] are considered as potential solutions to cope up with the problem of limited ground truth. As an example, Ryoo et al.[196] for interpreting brain perfusion SPECT trained a long short-term memory (LSTM) network on the unstructured text reports to predict abnormalities of basal perfusion and vascular reserve for each vascular territory. Using this trained LSTM model, the structured information of the text reports can be extracted. They developed a 3D CNN model with AUC 0.83 for basal perfusion and 0.89 for vascular reserve. This model can be utilized as a support system to identify the perfusion abnormalities and the quantitative scores of abnormalities.

### 6.4. Neuro-Symbolic AI models for detection and classification

AI techniques for detection and classification should not be trained only on the images, incorporating the complementary clinical information of the patients improves the performance and generalization of the developed AI models. Neuro-Symbolic AI models involve the integration of complementary approaches i.e. neural network (connectionist) and rule-based (symbolist) frameworks. Especially see Toosi et al. [197]for discussions of these two frameworks. Symbolic AI integrates contextual knowledge and generalizable rules into the AI framework. The rules that have been previously extracted (considering the decision made by physicians) can be used for feature prediction in new images. In the other words, these rules make the (deep) neural network biased toward the physicians' decision. Rule-based AI attracted significant attention in the early decades of AI research, but was later superseded by neural network (NN) approaches [189], and integration of the two appears to be in the horizon to significantly improve performance and generalizability of AI-based models.

Clinicians often use a set of conditions, different image modalities for a patient, treatment and surgery history, as well as biological and physiological conditions to decide about a suspicious lesion based on their experiences [198]. By incorporating these "rules" into the AI models for detection, classification and segmentation, we can move forward to neuro-symbolic neural networks and explainable AI techniques for detection and segmentation [199].

### 7. Conclusion

Considering the strength of AI techniques in performing effective image phenotyping including robust identification of patterns beyond visual assessments, there is significant potential for use in accelerating and automating detection and classification tasks in medical images and as an example oncologic PET imaging, as reviewed in the present work. AI techniques can also link radiomics signatures and biological properties extracted by radiologists and nuclear medicine physicians in routine clinical workflow. Having access to appropriate labeled data is vital for supervised AI techniques in derivation and validation of reliable detection and classification workflows. Alternatively, weakly- and un-supervised techniques along with transfer learning from different pathologies can be helpful in this regard. Considering the large existing database of electronic health records, applying NLP has shown potential to help tackle the labor-intensive task of labeling for supervised techniques. Besides, new insight into rule-based neural networks, i.e. neuro-symbolic networks, can obtain clues from decisions made by the physicians, with potential for further enhanced and generalizable AI clinical decision support systems.

**Acknowledgements**

This project was in part supported by the Natural Sciences and Engineering Research Council of Canada (NSERC) Discovery Grant RGPIN-2019-06467, and the Canadian Institutes of Health Research (CIHR) Project Grant PJT-173231.

References

1. Langlotz, C.P., *RadLex: a new method for indexing online educational materials*. 2006, Radiological Society of North America.2006.
2. Chernyak, V. and C.B. Sirlin, *LI-RADS: Future Directions.* Clinical Liver Disease, 2021. **17**(3): p. 149.
3. Bruno, M.A., E.A. Walker, and H.H. Abujudeh, *Understanding and confronting our mistakes: the epidemiology of error in radiology and strategies for error reduction.* Radiographics, 2015. **35**(6): p. 1668-1676.
4. Daniel, K., *Thinking, fast and slow*. 2017.2017.
5. Kim, Y.W. and L.T. Mansfield, *Fool me twice: delayed diagnoses in radiology with emphasis on perpetuated errors.* American journal of roentgenology, 2014. **202**(3): p. 465-470.
6. Zhaoping, L. and Z. Li, *Understanding vision: theory, models, and data*. 2014: Oxford University Press, USA.
7. Zhaoping, L., *A new framework for understanding vision from the perspective of the primary visual cortex.* Current Opinion in Neurobiology, 2019. **58**: p. 1-10.
8. Riesenhuber, M. and T. Poggio, *Hierarchical models of object recognition in cortex.* Nature neuroscience, 1999. **2**(11): p. 1019-1025.
9. DiCarlo, J.J. and D.D. Cox, *Untangling invariant object recognition.* Trends in cognitive sciences, 2007. **11**(8): p. 333-341.
10. DiCarlo, J.J., D. Zoccolan, and N.C. Rust, *How does the brain solve visual object recognition?* Neuron, 2012. **73**(3): p. 415-434.
11. Bar, M., *A cortical mechanism for triggering top-down facilitation in visual object recognition.* Journal of cognitive neuroscience, 2003. **15**(4): p. 600-609.
12. Kim, J.U., S.T. Kim, E.S. Kim, et al. *Towards high-performance object detection: Task-specific design considering classification and localization separation.* in *ICASSP 2020-2020 IEEE International Conference on Acoustics, Speech and Signal Processing (ICASSP)*. 2020. IEEE.
13. Murray, N.M., M. Unberath, G.D. Hager, et al., *Artificial intelligence to diagnose ischemic stroke and identify large vessel occlusions: a systematic review.* Journal of neurointerventional surgery, 2020. **12**(2): p. 156-164.
14. Chamberlin, J., M.R. Kocher, J. Waltz, et al., *Automated detection of lung nodules and coronary artery calcium using artificial intelligence on low-dose CT scans for lung cancer screening: accuracy and prognostic value.* BMC medicine, 2021. **19**(1): p. 1-14.

15. Kim, H.-E., H.H. Kim, B.-K. Han, et al., *Changes in cancer detection and false-positive recall in mammography using artificial intelligence: a retrospective, multireader study.* The Lancet Digital Health, 2020. **2**(3): p. e138-e148.
16. Winkel, D.J., T. Heye, T.J. Weikert, et al., *Evaluation of an AI-based detection software for acute findings in abdominal computed tomography scans: toward an automated work list prioritization of routine CT examinations.* Investigative radiology, 2019. **54**(1): p. 55-59.
17. Food, U. and D. Administration, *Computer-assisted detection devices applied to radiology images and radiology device data—Premarket notification [510 (k)] submissions.* Silver Spring: Food and Drug Administration, 2012.
18. Food, U. and D. Administration, *Clinical performance assessment: Considerations for computer-assisted detection devices applied to radiology images and radiology device data—Premarket approval (PMA) and premarket notification [510 (k)] submissions.* Silver Spring (MD/USA):[sn]. Disponível em:< https://www. fda. gov/media/77642/download, 2020.
19. Zhou, S.K., D. Rueckert, and G. Fichtinger, *Handbook of medical image computing and computer assisted intervention*. 2019: Academic Press.
20. Cho, J.S., S. Shrestha, N. Kagiyama, et al., *A network-based "phenomics" approach for discovering patient subtypes from high-throughput cardiac imaging data.* JACC: Cardiovascular Imaging, 2020. **13**(8): p. 1655-1670.
21. Houle, D., D.R. Govindaraju, and S. Omholt, *Phenomics: the next challenge.* Nature reviews genetics, 2010. **11**(12): p. 855-866.
22. Hoyles, L., J.-M. Fernandez-Real, M. Federici, et al., *Molecular phenomics and metagenomics of hepatic steatosis in non-diabetic obese women.* Nature medicine, 2018. **24**(7): p. 1070-1080.
23. Zbuk, K.M. and C. Eng, *Cancer phenomics: RET and PTEN as illustrative models.* Nature Reviews Cancer, 2007. **7**(1): p. 35-45.
24. Bizhanova, A. and P. Kopp, *Genetics and phenomics of Pendred syndrome.* Molecular and cellular endocrinology, 2010. **322**(1-2): p. 83-90.
25. Bourdais, G., P. Burdiak, A. Gauthier, et al., *Large-scale phenomics identifies primary and fine-tuning roles for CRKs in responses related to oxidative stress.* PLoS Genetics, 2015. **11**(7): p. e1005373.
26. Kafkas, Ş., S. Althubaiti, G.V. Gkoutos, et al., *Linking common human diseases to their phenotypes; development of a resource for human phenomics.* Journal of biomedical semantics, 2021. **12**(1): p. 1-15.
27. Seifert, R., M. Weber, E. Kocakavuk, et al., *Artificial Intelligence and Machine Learning in Nuclear Medicine: Future Perspectives.* Seminars in Nuclear Medicine, 2021. **51**(2): p. 170-177.
28. Hatt, M., F. Tixier, D. Visvikis, et al., *Radiomics in PET/CT: more than meets the eye?* Journal of Nuclear Medicine, 2017. **58**(3): p. 365-366.
29. Orlhac, F., C. Nioche, I. Klyuzhin, et al., *Radiomics in PET imaging: a practical guide for newcomers.* PET Clinics, 2021.
30. Horvat, N., D.D. Bates, and I. Petkovska, *Novel imaging techniques of rectal cancer: what do radiomics and radiogenomics have to offer? A literature review.* Abdominal Radiology, 2019. **44**(11): p. 3764-3774.

31. Langlotz, C.P., B. Allen, B.J. Erickson, et al., *A roadmap for foundational research on artificial intelligence in medical imaging: from the 2018 NIH/RSNA/ACR/The Academy Workshop.* Radiology, 2019. **291**(3): p. 781-791.
32. Al-shamasneh, A.R.M. and U.H.B. Obaidellah, *Artificial intelligence techniques for cancer detection and classification: review study.* European Scientific Journal, 2017. **13**(3): p. 342-370.
33. D'Amore, B., S. Smolinski-Zhao, D. Daye, et al., *Role of Machine Learning and Artificial Intelligence in Interventional Oncology.* Current Oncology Reports, 2021. **23**(6): p. 1-8.
34. Visvikis, D., C.C. Le Rest, V. Jaouen, et al., *Artificial intelligence, machine (deep) learning and radio (geno) mics: definitions and nuclear medicine imaging applications.* European journal of nuclear medicine and molecular imaging, 2019. **46**(13): p. 2630-2637.
35. Oquab, M., L. Bottou, I. Laptev, et al. *Is object localization for free?-weakly-supervised learning with convolutional neural networks.* in *Proceedings of the IEEE conference on computer vision and pattern recognition.* 2015.
36. Zhou, B., A. Khosla, A. Lapedriza, et al. *Learning deep features for discriminative localization.* in *Proceedings of the IEEE conference on computer vision and pattern recognition.* 2016.
37. Gillies, R.J., P.E. Kinahan, and H. Hricak, *Radiomics: Images Are More than Pictures, They Are Data.* Radiology, 2016. **278**(2): p. 563-577.
38. El Naqa, I., P.W. Grigsby, A. Apte, et al., *Exploring feature-based approaches in PET images for predicting cancer treatment outcomes.* Pattern recognition, 2009. **42**(6): p. 1162-1171.
39. Vallieres, M., E. Kay-Rivest, L.J. Perrin, et al., *Radiomics strategies for risk assessment of tumour failure in head-and-neck cancer.* Scientific reports, 2017. **7**(1): p. 1-14.
40. Kidd, E.A., I. El Naqa, B.A. Siegel, et al., *FDG-PET-based prognostic nomograms for locally advanced cervical cancer.* Gynecologic oncology, 2012. **127**(1): p. 136-140.
41. Vallières, M., C.R. Freeman, S.R. Skamene, et al., *A radiomics model from joint FDG-PET and MRI texture features for the prediction of lung metastases in soft-tissue sarcomas of the extremities.* Physics in Medicine & Biology, 2015. **60**(14): p. 5471.
42. Vaidya, M., K.M. Creach, J. Frye, et al., *Combined PET/CT image characteristics for radiotherapy tumor response in lung cancer.* Radiotherapy and Oncology, 2012. **102**(2): p. 239-245.
43. Frood, R., C. Burton, C. Tsoumpas, et al., *Baseline PET/CT imaging parameters for prediction of treatment outcome in Hodgkin and diffuse large B cell lymphoma: a systematic review.* European Journal of Nuclear Medicine and Molecular Imaging, 2021: p. 1-23.
44. Parmar, C., P. Grossmann, J. Bussink, et al., *Machine learning methods for quantitative radiomic biomarkers.* Scientific reports, 2015. **5**(1): p. 1-11.
45. Currie, G. and E. Rohren. *Intelligent imaging in nuclear medicine: the principles of artificial intelligence, machine learning and deep learning.* in *Seminars in Nuclear Medicine.* 2020. Elsevier.
46. Kotrotsou, A., P.O. Zinn, and R.R. Colen, *Radiomics in brain tumors: an emerging technique for characterization of tumor environment.* Magnetic Resonance Imaging Clinics, 2016. **24**(4): p. 719-729.

47. Li, X., G. Yin, Y. Zhang, et al., *Predictive power of a radiomic signature based on 18F-FDG PET/CT images for EGFR mutational status in NSCLC.* Frontiers in oncology, 2019. **9**: p. 1062.
48. Lartizien, C., M. Rogez, E. Niaf, et al., *Computer-aided staging of lymphoma patients with FDG PET/CT imaging based on textural information.* IEEE journal of biomedical and health informatics, 2013. **18**(3): p. 946-955.
49. Avanzo, M., L. Wei, J. Stancanello, et al., *Machine and deep learning methods for radiomics.* Medical physics, 2020. **47**(5): p. e185-e202.
50. Kebir, S., Z. Khurshid, F.C. Gaertner, et al., *Unsupervised consensus cluster analysis of [18F]-fluoroethyl-L-tyrosine positron emission tomography identified textural features for the diagnosis of pseudoprogression in high-grade glioma.* Oncotarget, 2017. **8**(5): p. 8294.
51. Yousefirizi, F., A.K. Jha, J. Brosch-Lenz, et al., *Toward High-Throughput Artificial Intelligence-Based Segmentation in Oncological PET Imaging.* PET Clinics, 2021. **16**(4): p. 577-596.
52. Afshar, P., A. Mohammadi, K.N. Plataniotis, et al., *From handcrafted to deep-learning-based cancer radiomics: challenges and opportunities.* IEEE Signal Processing Magazine, 2019. **36**(4): p. 132-160.
53. Van Griethuysen, J.J., A. Fedorov, C. Parmar, et al., *Computational radiomics system to decode the radiographic phenotype.* Cancer research, 2017. **77**(21): p. e104-e107.
54. Nioche, C., F. Orlhac, S. Boughdad, et al., *LIFEx: a freeware for radiomic feature calculation in multimodality imaging to accelerate advances in the characterization of tumor heterogeneity.* Cancer research, 2018. **78**(16): p. 4786-4789.
55. Ashrafinia, S., M. DiGianvittorio, S. Rowe, et al., *Reproducibility and reliability of radiomic features in 18F-DCFPyL PET/CT imaging of prostate cancer.* Journal of Nuclear Medicine, 2017. **58**(supplement 1): p. 503-503.
56. Deasy, J.O., A.I. Blanco, and V.H. Clark, *CERR: a computational environment for radiotherapy research.* Medical physics, 2003. **30**(5): p. 979-985.
57. Zwanenburg, A., M. Vallières, M.A. Abdalah, et al., *The image biomarker standardization initiative: standardized quantitative radiomics for high-throughput image-based phenotyping.* Radiology, 2020. **295**(2): p. 328-338.
58. Castiglioni, I., L. Rundo, M. Codari, et al., *AI applications to medical images: From machine learning to deep learning.* Physica Medica, 2021. **83**: p. 9-24.
59. Ashrafinia, S., P. Dalaie, M.S. Sadaghiani, et al., *Radiomics Analysis of Clinical Myocardial Perfusion Stress SPECT Images to Identify Coronary Artery Calcification.* medRxiv, 2021.
60. Zhang, Y., A. Oikonomou, A. Wong, et al., *Radiomics-based prognosis analysis for non-small cell lung cancer.* Scientific reports, 2017. **7**(1): p. 1-8.
61. Rizzo, S., F. Botta, S. Raimondi, et al., *Radiomics: the facts and the challenges of image analysis.* European radiology experimental, 2018. **2**(1): p. 1-8.
62. Lohmann, P., K. Bousabarah, M. Hoevels, et al., *Radiomics in radiation oncology—basics, methods, and limitations.* Strahlentherapie und Onkologie, 2020: p. 1-8.
63. Oktay, O., J. Schlemper, L.L. Folgoc, et al., *Attention u-net: Learning where to look for the pancreas.* arXiv preprint arXiv:1804.03999, 2018.

64. Khalvati, F., J. Zhang, A.G. Chung, et al., *MPCaD: a multi-scale radiomics-driven framework for automated prostate cancer localization and detection.* BMC medical imaging, 2018. **18**(1): p. 1-14.
65. Geirhos, R., P. Rubisch, C. Michaelis, et al., *ImageNet-trained CNNs are biased towards texture; increasing shape bias improves accuracy and robustness.* arXiv preprint arXiv:1811.12231, 2018.
66. Islam, M.A., M. Kowal, P. Esser, et al., *Shape or texture: Understanding discriminative features in CNNs.* arXiv preprint arXiv:2101.11604, 2021.
67. Klyuzhin, I.S., Y. Xu, A. Ortiz, et al., *Testing the Ability of Convolutional Neural Networks to Learn Radiomic Features.* medRxiv, 2020.
68. Kim, J., S. Seo, S. Ashrafinia, et al., *Training of deep convolutional neural nets to extract radiomic signatures of tumors.* Journal of Nuclear Medicine, 2019. **60**(supplement 1): p. 406-406.
69. Bizzego, A., N. Bussola, D. Salvalai, et al. *Integrating deep and radiomics features in cancer bioimaging.* in *2019 IEEE Conference on Computational Intelligence in Bioinformatics and Computational Biology (CIBCB)*. 2019. IEEE.
70. Peng, H., D. Dong, M.-J. Fang, et al., *Prognostic value of deep learning PET/CT-based radiomics: potential role for future individual induction chemotherapy in advanced nasopharyngeal carcinoma.* Clinical Cancer Research, 2019. **25**(14): p. 4271-4279.
71. Rizzo, A., E.K.A. Triumbari, R. Gatta, et al., *The role of 18F-FDG PET/CT radiomics in lymphoma.* Clinical and Translational Imaging, 2021: p. 1-10.
72. Domingues, I., G. Pereira, P. Martins, et al., *Using deep learning techniques in medical imaging: a systematic review of applications on CT and PET.* Artificial Intelligence Review, 2020. **53**(6): p. 4093-4160.
73. Weisman, A.J., M.W. Kieler, S.B. Perlman, et al., *Convolutional Neural Networks for Automated PET/CT Detection of Diseased Lymph Node Burden in Patients with Lymphoma.* Radiology: Artificial Intelligence, 2020. **2**(5): p. e200016.
74. Gruetzemacher, R., A. Gupta, and D. Paradice, *3D deep learning for detecting pulmonary nodules in CT scans.* Journal of the American Medical Informatics Association, 2018. **25**(10): p. 1301-1310.
75. Barbu, A., M. Suehling, X. Xu, et al., *Automatic detection and segmentation of lymph nodes from CT data.* IEEE Transactions on Medical Imaging, 2011. **31**(2): p. 240-250.
76. Cherry, K.M., S. Wang, E.B. Turkbey, et al. *Abdominal lymphadenopathy detection using random forest.* in *Medical Imaging 2014: Computer-Aided Diagnosis*. 2014. International Society for Optics and Photonics.
77. Topol, E.J., *High-performance medicine: the convergence of human and artificial intelligence.* Nature medicine, 2019. **25**(1): p. 44-56.
78. Gaonkar, B., J. Beckett, M. Attiah, et al., *Eigenrank by committee: Von-Neumann entropy based data subset selection and failure prediction for deep learning based medical image segmentation.* Medical Image Analysis, 2021. **67**: p. 101834.
79. Karimi, D., H. Dou, S.K. Warfield, et al., *Deep learning with noisy labels: Exploring techniques and remedies in medical image analysis.* Medical Image Analysis, 2020. **65**: p. 101759.
80. Shin, H.-C., H.R. Roth, M. Gao, et al., *Deep convolutional neural networks for computer-aided detection: CNN architectures, dataset characteristics and transfer learning.* IEEE transactions on medical imaging, 2016. **35**(5): p. 1285-1298.

81. Cai, J., A.P. Harrison, Y. Zheng, et al., *Lesion-harvester: Iteratively mining unlabeled lesions and hard-negative examples at scale.* IEEE Transactions on Medical Imaging, 2020. **40**(1): p. 59-70.
82. Huang, X., W. Sun, T.-L.B. Tseng, et al., *Fast and fully-automated detection and segmentation of pulmonary nodules in thoracic CT scans using deep convolutional neural networks.* Computerized Medical Imaging and Graphics, 2019. **74**: p. 25-36.
83. Zhu, W., C. Liu, W. Fan, et al. *Deeplung: Deep 3d dual path nets for automated pulmonary nodule detection and classification.* in *2018 IEEE Winter Conference on Applications of Computer Vision (WACV).* 2018. IEEE.
84. Xie, Z., *3D Region Proposal U-Net with Dense and Residual Learning for Lung Nodule Detection.* LUNA16, 2017.
85. George, J., S. Skaria, and V. Varun. *Using YOLO based deep learning network for real time detection and localization of lung nodules from low dose CT scans.* in *Medical Imaging 2018: Computer-Aided Diagnosis.* 2018. International Society for Optics and Photonics.
86. Wang, D., Y. Zhang, K. Zhang, et al. *Focalmix: Semi-supervised learning for 3d medical image detection.* in *Proceedings of the IEEE/CVF Conference on Computer Vision and Pattern Recognition.* 2020.
87. Jiang, H., H. Ma, W. Qian, et al., *An automatic detection system of lung nodule based on multigroup patch-based deep learning network.* IEEE journal of biomedical and health informatics, 2017. **22**(4): p. 1227-1237.
88. Huang, X., J. Shan, and V. Vaidya. *Lung nodule detection in CT using 3D convolutional neural networks.* in *2017 IEEE 14th International Symposium on Biomedical Imaging (ISBI 2017).* 2017. IEEE.
89. Ding, J., A. Li, Z. Hu, et al. *Accurate pulmonary nodule detection in computed tomography images using deep convolutional neural networks.* in *International Conference on Medical Image Computing and Computer-Assisted Intervention.* 2017. Springer.
90. Vivanti, R., A. Szeskin, N. Lev-Cohain, et al., *Automatic detection of new tumors and tumor burden evaluation in longitudinal liver CT scan studies.* International journal of computer assisted radiology and surgery, 2017. **12**(11): p. 1945-1957.
91. Ghesu, F.-C., B. Georgescu, Y. Zheng, et al., *Multi-scale deep reinforcement learning for real-time 3D-landmark detection in CT scans.* IEEE transactions on pattern analysis and machine intelligence, 2017. **41**(1): p. 176-189.
92. Xie, H., D. Yang, N. Sun, et al., *Automated pulmonary nodule detection in CT images using deep convolutional neural networks.* Pattern Recognition, 2019. **85**: p. 109-119.
93. DENG, C.-L., H.-Y. JIANG, and H.-M. LI, *Automated high uptake regions recognition and lymphoma detection based on fully convolutional networks on chest and abdomen PET image.* DEStech Transactions on Biology and Health, 2017(icmsb).
94. Afshari, S., A. BenTaieb, and G. Hamarneh, *Automatic localization of normal active organs in 3D PET scans.* Computerized Medical Imaging and Graphics, 2018. **70**: p. 111-118.
95. Kawakami, M., H. Sugimori, K. Hirata, et al., *Evaluation of Automatic Detection of Abnormal Uptake by Deep Learning and Combination Technique in FDG-PET Images.* 2020, Soc Nuclear Med.2020.

96. Schwyzer, M., D.A. Ferraro, U.J. Muehlematter, et al., *Automated detection of lung cancer at ultralow dose PET/CT by deep neural networks–initial results.* Lung Cancer, 2018. **126**: p. 170-173.
97. Blanc-Durand, P., A. Van Der Gucht, N. Schaefer, et al., *Automatic lesion detection and segmentation of 18F-FET PET in gliomas: a full 3D U-Net convolutional neural network study.* PLoS One, 2018. **13**(4): p. e0195798.
98. Bi, L., J. Kim, A. Kumar, et al., *Automatic detection and classification of regions of FDG uptake in whole-body PET-CT lymphoma studies.* Computerized Medical Imaging and Graphics, 2017. **60**: p. 3-10.
99. Xu, L., G. Tetteh, J. Lipkova, et al., *Automated whole-body bone lesion detection for multiple myeloma on 68Ga-Pentixafor PET/CT imaging using deep learning methods.* Contrast media & molecular imaging, 2018. **2018**.
100. Teramoto, A., H. Fujita, O. Yamamuro, et al., *Automated detection of pulmonary nodules in PET/CT images: Ensemble false-positive reduction using a convolutional neural network technique.* Medical physics, 2016. **43**(6Part1): p. 2821-2827.
101. Kumar, A., M. Fulham, D. Feng, et al., *Co-learning feature fusion maps from PET-CT images of lung cancer.* IEEE Transactions on Medical Imaging, 2019. **39**(1): p. 204-217.
102. Borrelli, P., J. Ly, R. Kaboteh, et al., *AI-based detection of lung lesions in [18 F] FDG PET-CT from lung cancer patients.* EJNMMI physics, 2021. **8**(1): p. 1-11.
103. Weisman, A., M. Kieler, S. Perlman, et al., *Ensemble 3D Convolutional Neural Networks for Automated Detection of Diseased Lymph Nodes*. 2020, Soc Nuclear Med.2020.
104. Punithavathy, K., S. Poobal, and M. Ramya, *Performance evaluation of machine learning techniques in lung cancer classification from PET/CT images.* FME Transactions, 2019. **47**(3): p. 418-423.
105. Zhao, Y., A. Gafita, B. Vollnberg, et al., *Deep neural network for automatic characterization of lesions on 68 Ga-PSMA-11 PET/CT.* European journal of nuclear medicine and molecular imaging, 2020. **47**(3): p. 603-613.
106. Zhou, Y., J. Yu, M. Liu, et al., *A machine learning-based parametric imaging algorithm for noninvasive quantification of dynamic [68Ga] DOTATATE PET-CT*. 2019, Soc Nuclear Med.2019.
107. Vauclin, S., K. Doyeux, S. Hapdey, et al., *Development of a generic thresholding algorithm for the delineation of 18FDG-PET-positive tissue: application to the comparison of three thresholding models.* Physics in Medicine & Biology, 2009. **54**(22): p. 6901.
108. Hellwig, D., T.P. Graeter, D. Ukena, et al., *18F-FDG PET for mediastinal staging of lung cancer: which SUV threshold makes sense?* Journal of Nuclear Medicine, 2007. **48**(11): p. 1761-1766.
109. Wahl, R.L., H. Jacene, Y. Kasamon, et al., *From RECIST to PERCIST: evolving considerations for PET response criteria in solid tumors.* Journal of nuclear medicine, 2009. **50**(Suppl 1): p. 122S-150S.
110. Kawakami, M., K. Hirata, S. Furuya, et al., *Development of combination methods for detecting malignant uptakes based on physiological uptake detection using object detection with PET-CT MIP images.* Frontiers in Medicine, 2020. **7**.
111. Redmon, J., S. Divvala, R. Girshick, et al. *You only look once: Unified, real-time object detection*. in *Proceedings of the IEEE conference on computer vision and pattern recognition*. 2016.

112. Bi, L., J. Kim, A. Kumar, et al., *Synthesis of positron emission tomography (PET) images via multi-channel generative adversarial networks (GANs)*, in *molecular imaging, reconstruction and analysis of moving body organs, and stroke imaging and treatment*. 2017, Springer. p. 43-51.
113. Ben-Cohen, A., E. Klang, S.P. Raskin, et al., *Cross-modality synthesis from CT to PET using FCN and GAN networks for improved automated lesion detection.* Engineering Applications of Artificial Intelligence, 2019. **78**: p. 186-194.
114. Wei, L. and I. El Naqa. *AI for Response Evaluation With PET/CT*. in *Seminars in Nuclear Medicine*. 2020. Elsevier.
115. Roth, H.R., L. Lu, J. Liu, et al., *Improving computer-aided detection using convolutional neural networks and random view aggregation.* IEEE transactions on medical imaging, 2015. **35**(5): p. 1170-1181.
116. Afifi, A. and T. Nakaguchi. *Unsupervised detection of liver lesions in CT images*. in *2015 37th Annual International Conference of the IEEE Engineering in Medicine and Biology Society (EMBC)*. 2015. IEEE.
117. de Vos, B.D., J.M. Wolterink, P.A. de Jong, et al., *ConvNet-based localization of anatomical structures in 3-D medical images.* IEEE transactions on medical imaging, 2017. **36**(7): p. 1470-1481.
118. Shen, Y., N. Wu, J. Phang, et al., *An interpretable classifier for high-resolution breast cancer screening images utilizing weakly supervised localization.* Medical image analysis, 2021. **68**: p. 101908.
119. Feng, X., J. Yang, A.F. Laine, et al. *Discriminative localization in CNNs for weakly-supervised segmentation of pulmonary nodules*. in *International conference on medical image computing and computer-assisted intervention*. 2017. Springer.
120. Schlemper, J., O. Oktay, L. Chen, et al., *Attention-gated networks for improving ultrasound scan plane detection.* arXiv preprint arXiv:1804.05338, 2018.
121. Schwyzer, M., K. Martini, D.C. Benz, et al., *Artificial intelligence for detecting small FDG-positive lung nodules in digital PET/CT: impact of image reconstructions on diagnostic performance.* European radiology, 2020. **30**(4): p. 2031-2040.
122. Gu, D., G. Liu, and Z. Xue, *On the performance of lung nodule detection, segmentation and classification.* Computerized Medical Imaging and Graphics, 2021. **89**: p. 101886.
123. Edenbrandt, L., P. Borrelli, J. Ulen, et al., *Automated analysis of PSMA-PET/CT studies using convolutional neural networks.* medRxiv, 2021.
124. Lee, J.J., H. Yang, B.L. Franc, et al., *Deep learning detection of prostate cancer recurrence with 18 F-FACBC (fluciclovine, Axumin®) positron emission tomography.* European journal of nuclear medicine and molecular imaging, 2020. **47**(13): p. 2992-2997.
125. Polymeri, E., M. Sadik, R. Kaboteh, et al., *Deep learning-based quantification of PET/CT prostate gland uptake: association with overall survival.* Clinical physiology and functional imaging, 2020. **40**(2): p. 106-113.
126. Perk, T., T. Bradshaw, S. Chen, et al., *Automated classification of benign and malignant lesions in 18F-NaF PET/CT images using machine learning.* Physics in Medicine & Biology, 2018. **63**(22): p. 225019.
127. Rattan, R., T. Kataria, S. Banerjee, et al., *Artificial intelligence in oncology, its scope and future prospects with specific reference to radiation oncology.* BJR| Open, 2019. **1**(xxxx): p. 20180031.

128. Cheung, H. and D. Rubin, *Challenges and opportunities for artificial intelligence in oncological imaging.* Clinical Radiology, 2021.
129. Leung, K., M.S. Sadaghiani, P. Dalaie, et al., *A deep learning-based approach for lesion classification in 3D 18F-DCFPyL PSMA PET images of patients with prostate cancer*. 2020, Soc Nuclear Med.2020.
130. Teramoto, A., H. Fujita, K. Takahashi, et al., *Hybrid method for the detection of pulmonary nodules using positron emission tomography/computed tomography: a preliminary study.* International journal of computer assisted radiology and surgery, 2014. **9**(1): p. 59-69.
131. Sibille, L., R. Seifert, N. Avramovic, et al., *18F-FDG PET/CT uptake classification in lymphoma and lung cancer by using deep convolutional neural networks.* Radiology, 2020. **294**(2): p. 445-452.
132. Kawauchi, K., S. Furuya, K. Hirata, et al., *A convolutional neural network-based system to classify patients using FDG PET/CT examinations.* BMC cancer, 2020. **20**(1): p. 1-10.
133. Moitra, D. and R.K. Mandal, *Classification of non-small cell lung cancer using one-dimensional convolutional neural network.* Expert Systems with Applications, 2020. **159**: p. 113564.
134. Acar, E., A. Leblebici, B.E. Ellidokuz, et al., *Machine learning for differentiating metastatic and completely responded sclerotic bone lesion in prostate cancer: a retrospective radiomics study.* The British journal of radiology, 2019. **92**(1101): p. 20190286.
135. Grossiord, E., H. Talbot, N. Passat, et al. *Automated 3D lymphoma lesion segmentation from PET/CT characteristics*. in *2017 IEEE 14th international symposium on biomedical imaging (ISBI 2017)*. 2017. IEEE.
136. Capobianco, N., M. Meignan, A.-S. Cottereau, et al., *Deep-Learning 18F-FDG Uptake Classification Enables Total Metabolic Tumor Volume Estimation in Diffuse Large B-Cell Lymphoma.* Journal of Nuclear Medicine, 2021. **62**(1): p. 30-36.
137. Sibille, L., N. Avramovic, B. Spottiswoode, et al., *PET uptake classification in lymphoma and lung cancer using deep learning*. 2018, Soc Nuclear Med.2018.
138. Du, D., H. Feng, W. Lv, et al., *Machine learning methods for optimal radiomics-based differentiation between recurrence and inflammation: application to nasopharyngeal carcinoma post-therapy PET/CT images.* Molecular imaging and biology, 2020. **22**(3): p. 730-738.
139. Du, D., J. Gu, X. Chen, et al., *Integration of PET/CT radiomics and semantic features for differentiation between active pulmonary tuberculosis and lung cancer.* Molecular Imaging and Biology, 2021. **23**(2): p. 287-298.
140. Lawhn-Heath, C., A. Salavati, S.C. Behr, et al., *Prostate-specific membrane antigen PET in prostate cancer.* Radiology, 2021. **299**(2): p. 248-260.
141. Sartor, O., J. de Bono, K.N. Chi, et al., *Lutetium-177–PSMA-617 for Metastatic Castration-Resistant Prostate Cancer.* New England Journal of Medicine, 2021.
142. Brosch-Lenz, J., F. Yousefirizi, K. Zukotynski, et al., *Role of Artificial Intelligence in Theranostics: Toward Routine Personalized Radiopharmaceutical Therapies.* PET clinics, 2021. **16**(4): p. 627-641.
143. Leung, K., S. Ashrafinia, M.S. Sadaghiani, et al., *A fully automated deep-learning based method for lesion segmentation in 18F-DCFPyL PSMA PET images of patients with prostate cancer.* Journal of Nuclear Medicine, 2019. **60**(supplement 1): p. 399-399.

144. Oldenhuis, C., S. Oosting, J. Gietema, et al., *Prognostic versus predictive value of biomarkers in oncology.* European journal of cancer, 2008. **44**(7): p. 946-953.
145. Lambin, P., E. Roelofs, B. Reymen, et al., *Rapid Learning health care in oncology'–an approach towards decision support systems enabling customised radiotherapy.* Radiotherapy and Oncology, 2013. **109**(1): p. 159-164.
146. Martens, R.M., T. Koopman, D.P. Noij, et al., *Predictive value of quantitative 18 F-FDG-PET radiomics analysis in patients with head and neck squamous cell carcinoma.* EJNMMI research, 2020. **10**(1): p. 1-15.
147. Wang, X. and Z. Lu, *Radiomics Analysis of PET and CT Components of 18F-FDG PET/CT Imaging for Prediction of Progression-Free Survival in Advanced High-Grade Serous Ovarian Cancer.* Frontiers in oncology, 2021. **11**.
148. Lv, W., S. Ashrafinia, J. Ma, et al., *Multi-level multi-modality fusion radiomics: application to PET and CT imaging for prognostication of head and neck cancer.* IEEE journal of biomedical and health informatics, 2019. **24**(8): p. 2268-2277.
149. Cottereau, A.-S., M. Meignan, C. Nioche, et al., *Risk stratification in diffuse large B-cell lymphoma using lesion dissemination and metabolic tumor burden calculated from baseline PET/CT.* Annals of Oncology, 2021. **32**(3): p. 404-411.
150. Desseroit, M.-C., D. Visvikis, F. Tixier, et al., *Development of a nomogram combining clinical staging with 18 F-FDG PET/CT image features in non-small-cell lung cancer stage I–III.* European journal of nuclear medicine and molecular imaging, 2016. **43**(8): p. 1477-1485.
151. Katzman, J.L., U. Shaham, A. Cloninger, et al., *DeepSurv: personalized treatment recommender system using a Cox proportional hazards deep neural network.* BMC medical research methodology, 2018. **18**(1): p. 1-12.
152. Desbordes, P., S. Ruan, R. Modzelewski, et al., *Predictive value of initial FDG-PET features for treatment response and survival in esophageal cancer patients treated with chemo-radiation therapy using a random forest classifier.* PLoS One, 2017. **12**(3): p. e0173208.
153. Hosny, A., C. Parmar, T.P. Coroller, et al., *Deep learning for lung cancer prognostication: a retrospective multi-cohort radiomics study.* PLoS medicine, 2018. **15**(11): p. e1002711.
154. Salmanpour, M.R., M. Shamsaei, and A. Rahmim, *Feature selection and machine learning methods for optimal identification and prediction of subtypes in Parkinson's disease.* Computer Methods and Programs in Biomedicine, 2021. **206**: p. 106131.
155. Paul, R., S.H. Hawkins, Y. Balagurunathan, et al., *Deep feature transfer learning in combination with traditional features predicts survival among patients with lung adenocarcinoma.* Tomography, 2016. **2**(4): p. 388-395.
156. Lao, J., Y. Chen, Z.-C. Li, et al., *A deep learning-based radiomics model for prediction of survival in glioblastoma multiforme.* Scientific reports, 2017. **7**(1): p. 1-8.
157. Ypsilantis, P.-P., M. Siddique, H.-M. Sohn, et al., *Predicting response to neoadjuvant chemotherapy with PET imaging using convolutional neural networks.* PloS one, 2015. **10**(9): p. e0137036.
158. Pereira, G., *Deep Learning techniques for the evaluation of response to treatment in Hogdkin Lymphoma.* 2018, Universidade de Coimbra.2018.
159. Capobianco, N., M.A. Meignan, A.-S. Cottereau, et al., *Deep learning FDG uptake classification enables total metabolic tumor volume estimation in diffuse large B-cell lymphoma.* Journal of Nuclear Medicine, 2020: p. jnumed. 120.242412.

160. Wang, H., Z. Zhou, Y. Li, et al., *Comparison of machine learning methods for classifying mediastinal lymph node metastasis of non-small cell lung cancer from 18 F-FDG PET/CT images.* EJNMMI research, 2017. **7**(1): p. 1-11.
161. Noortman, W.A., D. Vriens, C.H. Slump, et al., *Adding the temporal domain to PET radiomic features.* PloS one, 2020. **15**(9): p. e0239438.
162. Carvalho, S., R. Leijenaar, E. Troost, et al., *Early variation of FDG-PET radiomics features in NSCLC is related to overall survival-the "delta radiomics" concept.* Radiotherapy and Oncology, 2016. **118**: p. S20-S21.
163. Fave, X., L. Zhang, J. Yang, et al., *Delta-radiomics features for the prediction of patient outcomes in non–small cell lung cancer.* Scientific reports, 2017. **7**(1): p. 1-11.
164. Nasief, H., C. Zheng, D. Schott, et al., *A machine learning based delta-radiomics process for early prediction of treatment response of pancreatic cancer.* NPJ precision oncology, 2019. **3**(1): p. 1-10.
165. Chelala, L., R. Hossain, E.A. Kazerooni, et al., *Lung-RADS Version 1.1: Challenges and a Look Ahead, From the AJR Special Series on Radiology Reporting and Data Systems.* American Journal of Roentgenology, 2021. **216**(6): p. 1411-1422.
166. Mazzei, M.A., L. Di Giacomo, G. Bagnacci, et al., *Delta-radiomics and response to neoadjuvant treatment in locally advanced gastric cancer—a multicenter study of GIRCG (Italian Research Group for Gastric Cancer).* Quantitative Imaging in Medicine and Surgery, 2021. **11**(6): p. 2376.
167. Wang, L., Z. Gao, C. Li, et al., *Computed Tomography–Based Delta-Radiomics Analysis for Discriminating Radiation Pneumonitis in Patients With Esophageal Cancer After Radiation Therapy.* International Journal of Radiation Oncology* Biology* Physics, 2021.
168. Liu, Y., H. Shi, S. Huang, et al., *Early prediction of acute xerostomia during radiation therapy for nasopharyngeal cancer based on delta radiomics from CT images.* Quantitative imaging in medicine and surgery, 2019. **9**(7): p. 1288.
169. Sushentsev, N., L. Rundo, O. Blyuss, et al., *Comparative performance of MRI-derived PRECISE scores and delta-radiomics models for the prediction of prostate cancer progression in patients on active surveillance.* European Radiology, 2021: p. 1-10.
170. Peeken, J.C., R. Asadpour, K. Specht, et al., *MRI-based Delta-Radiomics predicts pathologic complete response in high-grade soft-tissue sarcoma patients treated with neoadjuvant therapy.* Radiotherapy and Oncology, 2021.
171. Shayesteh, S., M. Nazari, A. Salahshour, et al., *Treatment response prediction using MRI-based pre-, post-, and delta-radiomic features and machine learning algorithms in colorectal cancer.* Medical physics, 2021.
172. Nardone, V., A. Reginelli, C. Guida, et al., *Delta-radiomics increases multicentre reproducibility: a phantom study.* Medical Oncology, 2020. **37**(5): p. 1-7.
173. Jin, W., M. Fatehi, K. Abhishek, et al., *Artificial intelligence in glioma imaging: challenges and advances.* Journal of neural engineering, 2020. **17**(2): p. 021002.
174. He, J., S.L. Baxter, J. Xu, et al., *The practical implementation of artificial intelligence technologies in medicine.* Nature medicine, 2019. **25**(1): p. 30-36.
175. Zhang, Y., Q.V. Liao, and R.K. Bellamy. *Effect of confidence and explanation on accuracy and trust calibration in AI-assisted decision making.* in *Proceedings of the 2020 Conference on Fairness, Accountability, and Transparency.* 2020.

176. Jin, W., X. Li, and G. Hamarneh, *One Map Does Not Fit All: Evaluating Saliency Map Explanation on Multi-Modal Medical Images.* arXiv preprint arXiv:2107.05047, 2021.
177. Zaharchuk, G. and G. Davidzon. *Artificial Intelligence for Optimization and Interpretation of PET/CT and PET/MR Images.* in *Seminars in Nuclear Medicine*. 2020.
178. Arabi, H., A. AkhavanAllaf, A. Sanaat, et al., *The promise of artificial intelligence and deep learning in PET and SPECT imaging.* Physica Medica, 2021. **83**: p. 122-137.
179. Langlotz, C.P., *Will artificial intelligence replace radiologists?* 2019, Radiological Society of North America.2019.
180. Freeman, K., J. Geppert, C. Stinton, et al., *Use of artificial intelligence for image analysis in breast cancer screening programmes: systematic review of test accuracy.* bmj, 2021. **374**.
181. Hatt, M., C.C. Le Rest, N. Antonorsi, et al. *Radiomics in PET/CT: Current Status and Future AI-Based Evolutions*. in *Seminars in Nuclear Medicine*. 2020. Elsevier.
182. Shakir, H., Y. Deng, H. Rasheed, et al., *Radiomics based likelihood functions for cancer diagnosis.* Scientific reports, 2019. **9**(1): p. 1-10.
183. Papanikolaou, N., C. Matos, and D.M. Koh, *How to develop a meaningful radiomic signature for clinical use in oncologic patients.* Cancer Imaging, 2020. **20**(1): p. 1-10.
184. Krishna, R., Y. Zhu, O. Groth, et al., *Visual genome: Connecting language and vision using crowdsourced dense image annotations.* International journal of computer vision, 2017. **123**(1): p. 32-73.
185. Kumar, V., Y. Gu, S. Basu, et al., *Radiomics: the process and the challenges.* Magnetic resonance imaging, 2012. **30**(9): p. 1234-1248.
186. Kersting, D., M. Weber, L. Umutlu, et al., *Using a Lymphoma and Lung Cancer Trained Neural Network to Predict the Outcome for Breast Cancer on FDG PET/CT Data.* Nuklearmedizin, 2021. **60**(02): p. V74.
187. Capobianco, N., A. Gafita, G. Platsch, et al., *Transfer learning of AI-based uptake classification from 18F-FDG PET/CT to 68Ga-PSMA-11 PET/CT for whole-body tumor burden assessment.* Journal of Nuclear Medicine, 2020. **61**(supplement 1): p. 1411-1411.
188. Wu, J., X.-Y. Chen, H. Zhang, et al., *Hyperparameter optimization for machine learning models based on Bayesian optimization.* Journal of Electronic Science and Technology, 2019. **17**(1): p. 26-40.
189. Sundar, L.K.S., O. Muzik, I. Buvat, et al., *Potentials and caveats of AI in hybrid imaging.* Methods, 2021. **188**: p. 4-19.
190. (ESR), E.S.o.R., *ESR paper on structured reporting in radiology.* Insights into imaging, 2018. **9**: p. 1-7.
191. Lakhani, P., W. Kim, and C.P. Langlotz, *Automated extraction of critical test values and communications from unstructured radiology reports: an analysis of 9.3 million reports from 1990 to 2011.* Radiology, 2012. **265**(3): p. 809-818.
192. Lipton, Z.C., J. Berkowitz, and C. Elkan, *A critical review of recurrent neural networks for sequence learning.* arXiv preprint arXiv:1506.00019, 2015.
193. Folio, L.R., L.B. Machado, and A.J. Dwyer, *Multimedia-enhanced radiology reports: concept, components, and challenges.* RadioGraphics, 2018. **38**(2): p. 462-482.
194. Do, H., F. Farhadi, Z. Xu, et al., *AI radiomics in a monogenic autoimmune disease: deep learning of routine radiologist annotations correlated with pathologically verified lung findings.* Oak Brook, Ill: Radiological Society of North America, 2019.

195. Willemink, M.J., W.A. Koszek, C. Hardell, et al., *Preparing medical imaging data for machine learning.* Radiology, 2020. **295**(1): p. 4-15.
196. Ryoo, H.G., H. Choi, and D.S. Lee, *Deep learning-based interpretation of basal/acetazolamide brain perfusion SPECT leveraging unstructured reading reports.* European journal of nuclear medicine and molecular imaging, 2020. **47**(9): p. 2186-2196.
197. Toosi, A., A.G. Bottino, B. Saboury, et al., *A brief history of AI: how to prevent another winter (a critical review).* PET clinics, 2021. **16**(4): p. 449-469.
198. Manhaeve, R., S. Dumancic, A. Kimmig, et al., *Deepproblog: Neural probabilistic logic programming.* Advances in Neural Information Processing Systems, 2018. **31**: p. 3749-3759.
199. Došilović, F.K., M. Brčić, and N. Hlupić. *Explainable artificial intelligence: A survey.* in *2018 41st International convention on information and communication technology, electronics and microelectronics (MIPRO).* 2018. Opatija, Croatia: IEEE.